# On the Value of Distribution Tail in the Valuation of Travel Time Variability

Zhaoqi Zang [b, a], Richard Batley [c], Xiangdong Xu [a, *], David Z.W. Wang [b]

[a] College of Transportation Engineering, Tongji University, Shanghai, China
[b] School of Civil and Environmental Engineering, Nanyang Technological University, Singapore
[c] Institute for Transport Studies, University of Leeds, Leeds, United Kingdom

**Abstract**

Extensive empirical studies show that the long distribution tail of travel time and the corresponding unexpected delay can have much more serious consequences than expected or moderate delay. However, the unexpected delay due to the distribution tail of travel time has received limited attention in recent studies of the valuation of travel time variability. As a complement to current valuation research, this paper proposes the concept of the value of travel time distribution tail, which quantifies the value that travelers place on reducing the unexpected delay for hedging against travel time variability. Methodologically, we define the summation of all unexpected delays as the unreliability area to quantify travel time distribution tail and show that it is a key element of two well-defined measures accounting for unreliable aspects of travel time. We then formally derive the value of distribution tail, show that it is distinct from the more established value of reliability (VOR), and combine it and the VOR in an overall value of travel time variability (VOV). We prove theoretically that the VOV exhibits diminishing marginal benefit in terms of the traveler's punctuality requirements under a validity condition. This implies that it may be economically inefficient for travelers to blindly pursue a higher probability of not being late. We then proceed to develop the concept of the travel time variability ratio, which gives the implicit cost of the punctuality requirement imposed on any given trip. Numerical examples reveal that the cost of travel time distribution tail can account for more than 10% of the trip cost, such that its omission could introduce non-trivial bias into route choice models and transportation appraisal more generally.

---

* Corresponding author.
E-mail addresses: zhaoqi.zang@ntu.edu.sg (Z. Zang), R.P.Batley@its.leeds.ac.uk (R. Batley), xiangdongxu@tongji.edu.cn (X. Xu), wangzhiwei@ntu.edu.sg (D.Z.W Wang).

# 1. INTRODUCTION

In a road network, travel time can be variable due to the inherent supply-side (e.g., adverse weather conditions, and traffic incidents) and demand-side uncertainties (e.g., special events, and temporal factors) (van Lint *et al*., 2008; Chen and Zhou, 2010); this is the basis of the well-established concept of travel time variability. Empirical studies show that travelers value travel time variability almost as much as they value the mean travel time (Hollander, 2006; Asensio and Matas, 2008; Li *et al*., 2010; Carrion and Levinson, 2013; Khalili *et al*., 2022). To account for travel time variability within models of the road network, different variability costs may be assigned to different routes using different route choice criterion (e.g., Jackson and Jucker, 1982; Bell, 2000; Lo et al., 2006; Chen and Zhou, 2010; Nie, 2011; Xu et al., 2017, etc. See Zang *et al*. (2022b) for a summary), and these may affect travelers' behavior. Moreover, the benefits of reduced travel time variability may be non-trivial in magnitude (Fosgerau, 2017), such that there is a strong case for including them in the appraisal of publicly funded transportation project (de Jong and Bliemer, 2015; New Zealand Transport Agency, 2016; Organization for Economic Co-operation and Development (OECD), 2016).

The literature has developed a range of valuation methods to explicitly quantify the cost of travel time variability. However, it is arguably the case that this literature has focused on valuing travel times within the body of the distribution pertaining to high probability delays of moderate duration and devoted considerably less attention to travel times within the tail of the distribution pertaining to low probability delays of substantial duration. It is undoubtedly the case that these low probability events could have considerably more serious implications for travel times (and associated costs) (Odgaard *et al.*, 2005; van Lint *et al*., 2008; Franklin and Karlström, 2009; Sikka and Hanley, 2013). Section 1.1 to follow will use a verified behavioral assumption to formalize the distinction between these two types of events, referring to the former as reliable aspect of travel time emanating from "expected delays", and the latter as unreliable aspect emanating from "unexpected delays". On this basis, the contribution of this paper is to derive distinct valuations of both reliable aspect and unreliable aspects of travel time variability.

**1.1 Behavior Response to Travel Time Variability**

For travelers faced with travel time variability, a verified behavioral assumption is that beyond the mean travel time $\mu$, they will add a safety margin $\delta$ to mitigate for the probability of being late (Garver, 1968; Knight, 1974; Hall, 1983; Senna, 1994). In other words, the safety margin is a buffer time the traveler is willing to 'pay' (in time units) in excess of the mean travel time to ensure a specified trip reliability. Note that the summation of mean travel time and safety margin is also interpreted as travel time budget in Lo et al. (2006), which can conditionally correspond to percentile travel time (Wu and Nie, 2011; Nie, 2011).

In this paper, that specified trip reliability will be referred to as the punctuality requirement $\tau$ of a trip, meaning that the confidence level of punctual (on-time) arrival of the trip is $\tau$. Put differently, the punctuality requirement $\tau$ means that the probability of completing the trip should be no less than the predefined threshold $\tau$. Hence, the safety margin is subjective on the part of the traveler and its value depends on the traveler's $\tau$. To illustrate this, we present Figure 1, in which the upper panel presents the probability density function (PDF) of a random travel time $T$, and the lower panel presents the cumulative distribution function (CDF) of $T$. In this paper, we define delay as the difference between actual travel time and mean travel time, i.e., delay = actual travel time − mean travel time[1]. Then, the safety margin $\delta$ can be interpreted as the maximum delay expected (or accepted) by the traveler, thus we use the term "expected delay" to depict the delay brought about by an actual trip travel time in between the mean travel time and $\mu+\delta$ in the lower panel of Figure 1. In other words, the safety margin corresponds to the question "*based on historical experience, how much would I expect the maximum delay to be for my trip*?". But note from Figure 1 that this leaves a small probability that the traveler will experience travel time in excess of $\mu+\delta$, and thus experience an actual delay longer than he/she expects. In the lower panel of Figure 1, we use the term "unexpected delay" to depict the delay brought about by an actual trip travel time longer than $\mu+\delta$. Now, we are ready to present the following definition: for the traveler who is willing to pay a safety margin $\delta$ to mitigate for travel time variability, if the actual travel time of a trip is less than $\mu+\delta$, then the trip is considered *reliable* because the delay will be no longer than the traveler expects; by contrast, if the actual travel time of a trip exceeds $\mu+\delta$, then the trip is considered *unreliable* because the traveler could still experience an unexpected delay (i.e., the actual delay minus the expected delay $\delta$). Against this background, the present paper will make a distinction

---

[1] In traffic flow theory and related studies concerned especially with congestion, delay may be defined as the difference between the actual travel time and the free-flow travel time.

between what we define as:
- **Reliable aspect of travel time**: expected risk/duration of delay, as reflected by the safety margin planned by travelers to offset such reliability, and shown by the shaded area (horizontal grid lines) in the upper panel of Figure 1.
- **Unreliable aspect of travel time**: unexpected risk/duration of delay, as reflected by the tail of the travel time distribution beyond the safety margin, and shown by the shaded area (vertical grid lines) in the upper panel of Figure 1.

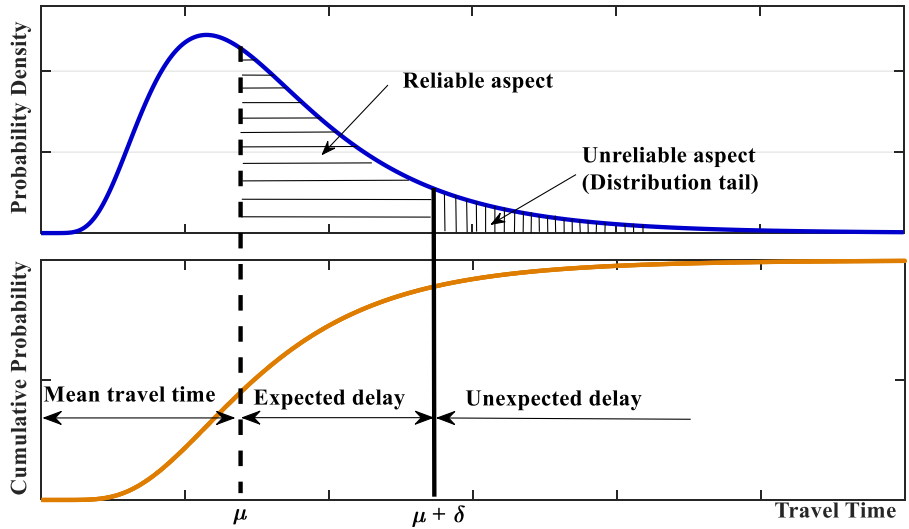

Figure 1. Illustration of reliable and unreliable aspects of travel time defined in this paper (modified from Xu *et al.* (2014))

Consequently, all travel time greater than $\mu+\delta$ in the right-hand tail of Figure 1 (i.e., the right-hand part of the PDF separated by the vertical solid line) is unreliable, meaning that travelers will experience *unexpected delay* even if they have allowed a safety margin $\delta$ to absorb the expected delay. Such *unexpected delay* resulting from the long distribution tail of travel time has received limited attention in previous studies of the cost of travel time variability. Instead, these studies focus mainly on the willingness-to-pay of travelers to improve travel time reliability (as opposed to unreliable aspect). Among these studies, the standard deviation and variance of travel time are the two most frequently used valuation measures (Hollander, 2006; Fosgerau and Karlström, 2010; Fosgerau and Engelson, 2011; Jenelius *et al.*, 2011; Jenelius, 2012; Uchida, 2014; Coulombel and de Palma, 2014; Fosgerau, 2017; Zhu *et al.*, 2018), while some contributions also use percentiles or ranges between percentiles (e.g., Lam and Small, 2001; Brownstone and Small, 2005; Small *et al.*, 2005). Interested readers may refer to Bates *et al.* (2001), Li *et al.* (2010), Carrion and Levinson (2013), Shams *et al.* (2017), and Zang *et al*. (2022b) for comprehensive reviews of theories and evidence. The standard deviation and variance may not however capture the distribution tail of travel time especially well, as they are measures of symmetric variables and can only quantify the extent of variation or dispersion from the mean value. Similarly, in considering the punctuality requirements $\tau$, the percentiles tend to focus on the mass of the travel time distribution, but omits consideration of the tail (Lee *et al*., 2019). It should be noted that the value of reliability defined in Fosgerau and Karlström (2010) could inexplicitly consider the impact of distribution tail (i.e., a part of sources resulted in the defined value of reliability) as its expression has included the mean lateness factor which depends on the shape of the standardized travel time distribution. However, this paper explicitly considers the distribution tail and its impact on traveler's departure time and then clearly quantifies the cost resulted from distribution tail only, to be introduced later.

Although unreliable aspect of travel time (as just defined) is not explicitly considered in conventional valuation studies of travel time variability, considerable empirical evidence verifies the existence of highly-skewed travel time distributions with long/fat tails (van Lint and van Zuylen, 2005; Fosgerau and Fukuda, 2012; Susilawati *et al*., 2013; Delhome *et al*., 2015; Kim and Mahmassani, 2015; Taylor, 2017; Zang *et al*., 2018b; Li, 2019). This could represent a notable omission from valuation studies, since unexpected delays that travelers experience due to the distribution tail could have much more serious consequences than expected delays. For example, among drivers unlucky enough to experience delay when traveling on dense freeway corridors in the Netherlands, the delay that 5% of the "unluckiest drivers" incurred was almost five times that of 50% of the "unlucky drivers" (van Lint *et al*., 2008). To avoid unexpected delays, Sikka and Hanley (2013) found that travelers would be willing to pay nearly half of the cost of the usual travel time (i.e., the summation of the mean travel time and expected delay). Franklin and Karlström (2009) and Odgaard *et al.* (2005) suggested that unexpected delays should be included in project appraisals to account for these more substantially delayed trips. The effects of the above-described notion of unexpected delay, associated with the long tail of the travel time distribution, have previously been considered in studies of network performance assessment (Xu *et al*., 2014, 2021), hazardous materials routing (Toumazis and Kwon, 2013, 2016; Su and Kwon, 2020), and route guidance systems (Lee *et al*., 2019). In summary, it is necessary to quantify and include the cost of travel time distribution tail in the valuation of travel time variability.

## 1.2 Main Contributions of This Paper

To complement existing valuation studies of travel time variability, this paper proposes the concept of the *value of travel time distribution tail* to quantify in monetary terms the unreliable aspects (i.e., unexpected delays) due to the long tail of the travel time distribution. Specifically, we define the summation of all unexpected delays as the unreliability area to quantify the distribution tail and show that the unreliability area is a key element of two existing well-defined measures, namely the mean-excess travel time (Chen and Zhou, 2010) and the reliability premium (Batley, 2007). We prove theoretically that there exists a proportional relationship between these two measures and find that the only difference between these two measures is the "weight" attached to the unreliability area. We then derive formulations of both the *value of distribution tail* (VODT) and *value of travel time reliability* (VOR), associated with the scheduling of a trip under uncertainty in travel time (Noland and Small, 1995; Small, 1982). To fully capture both the reliable and unreliable aspects of travel time, the VODT and VOR are then aggregated into what we define as the *value of travel time variability* (VOV). The VOV can be interpreted as the value of the additional time (safety margin) a traveler pays for hedging against travel time variability. Furthermore, we prove theoretically that the VOV has a diminishing marginal effect under a validity condition as the punctuality requirements imposed by travelers become more stringent. Numerical examples show that this validity condition is relatively slack and can be easily satisfied by empirical datasets. The diminishing marginal effect means that it may be economically inefficient for travelers to blindly pursue a higher probability of not being late. To this end, we develop the concept of a travel time variability ratio, based on the diminishing marginal effect of the VOV, to help travelers understand the implicit cost of the punctuality requirements they impose. Numerical examples show that the travel time variability ratio can help travelers to improve the balance between trip reliability and trip cost. In particular, these examples reveal that the cost of travel time distribution tail can account for more than 10% of the trip cost, such that its omission could introduce non-trivial bias into route choice models and transportation appraisal more generally.

In summary, the contributions of this paper are threefold. (1) To quantify the unexpected delays due to the long tail of the distribution (i.e., unreliable aspects) of travel time in monetary units, we propose the concept of VODT and derive its formulation. (2) We integrate the VODT and VOR into the VOV to fully quantify the costs of both the reliable and unreliable aspects of travel time and then theoretically prove the diminishing marginal effect of the VOV. (3) Based on this diminishing marginal effect, we further develop the travel time variability ratio to reveal that travelers must accept the diminishing effect of the VOV in order to improve their trip reliability. Accordingly, the travel time variability ratio can help travelers to balance their punctuality requirements against trip cost. It should be noted that the purpose of this paper is not to claim that the proposed valuation method outperforms current methods of valuing reliability. Instead, the proposed method is presented as a complement to current valuation methods, through quantifying the value of travel time distribution tail that is additional to the value of reliability.

The layout of this paper is as follows. The measures of quantifying reliable and unreliable aspect of travel time and derivations of the VODT, VOR, and VOV are introduced and discussed in Section 2. In Section 3, we illustrate the necessity of considering the unreliable aspect of travel time, test the feasibility of the validity condition in propositions, and illustrate how the travel time variability ratio works. Finally, Section 4 summarizes and concludes the work. Detailed derivations and proofs of most propositions and lemmas are given in the Appendixes. The notation used is listed at the end of the paper.

## 2. METHODOLOGY

In this section, we first introduce the scheduling model, followed by the measures used to include unreliable aspect of travel time within this model. Finally, we derive and discuss the VODT, VOR, VOV, and travel time variability ratio.

### 2.1 The Schedule Delay Model and Unreliability Area

In the ($\alpha$, $\beta$, $\gamma$) scheduling model (Small, 1982; Noland and Small, 1995), the individual traveler holds a preference for being early or late relative to a preferred arrival time (PAT). Let $A$, $D$, and $T$ denote the arrival time, departure time, and stochastic travel time, respectively. The utility of a trip for a PAT in terms of the step model is given by

$$U(D,T) = -\left\{\alpha T + \beta(PAT - A)^+ + \gamma(A - PAT)^+\right\} \tag{1}$$

where $(.)^+$ denotes a function such that $x^+ = x$ if $x > 0$, and 0 otherwise; $(PAT - A)^+$ is schedule delay early (SDE); and $(A - PAT)^+$ is schedule delay late (SDL). The traveler's preference parameters $\alpha$, $\beta$, and $\gamma$ are all positive, which represent the marginal utilities of mean travel time, SDE and SDL, respectively. Since the function as a whole is negative, the traveler will maximize his/her utility by minimizing $T$, SDE and SDL.

Without loss of generality, an individual's PAT can be normalized to time zero, i.e., *PAT* = 0. Normalizing an individual's PAT to be zero, i.e., *PAT* = 0, is a trick developed by Fosgerau and Karlström (2010) to simplify mathematical deductions. Referring to the Eq. (1), we only focus on the difference between *PAT* and arrival time, regardless of the specified value of *PAT*. As a result, $D < 0$ because the travel time must be positive for a trip under certainty. Then, for a traveler, (a) if travel time $T < -D$, he/she will be early; (b) if $T > -D$, he/she will be late; and (c) if $T = -D$, he/she will arrive at exactly the PAT. On this basis, Eq. (1) can be simplified to Eq. (2) (see Appendix A for details).

$$U(D,T) = -\left\{(\alpha-\beta)T + (\beta+\gamma)(T+D)\big|_{A \geq 0} - \beta D\right\} \quad (2)$$

For a trip under travel time certainty, the utility of early and late arrivals is given by

$$U(D,T) = \begin{cases} -\{(\alpha-\beta)T - \beta D\} &, A < 0 \\ -\{(\alpha-\beta)T + (\beta+\gamma)(T+D) - \beta D\} &, A \geq 0 \end{cases} \quad (3)$$

However, more realistic is some level of travel time variability due to demand-side or supply-side factors. Then, for a trip under travel time uncertainty, the expected utility becomes (Fosgerau and Karlström, 2010; Zang *et al.*, 2018a; Li, 2019)

$$EU(D,T) = -\left\{(\alpha-\beta)\mu - \beta D + (\beta+\gamma)\left(\int_{\frac{-D-\mu}{\sigma}}^{\infty}(\mu+\sigma x + D)\cdot f(x)dx\right)\right\} \quad (4)$$

where the standardized travel time $X = (T-\mu)/\sigma$; and $\mu$ and $\sigma$ represent the mean and standard deviation of $T$, respectively. Besides, $X$ has a PDF $f(x)$ and CDF $F(x)$. Note that for a trip under uncertainty, $D$ can be both negative and positive.

To maximize the expected utility of a trip under uncertainty, the optimal departure time $D^*$ for travelers satisfies the first-order condition of Eq. (4):

$$D^* = -\mu - \sigma F^{-1}\left(\frac{\gamma}{\beta+\gamma}\right) \quad (5)$$

As shown in Eq. (5), $D^*$ only considers the reliable aspect of travel time via a specified safety margin (i.e., $\sigma F^{-1}(\gamma/(\gamma+\beta))$). However, $D^*$ does not consider the unreliable aspect of travel time shown by the shaded area in Figure 2, i.e., any travel time in the distribution tail longer than $\mu+\sigma F^{-1}(\tau)$, where $\tau = \gamma/(\gamma+\beta)^2$. Figure 2 presents the inverse CDF (i.e., percentile function) of travel time. The shaded area in Figure 2 is defined as the *unreliability area* $S_u$, which includes all trips with a travel time longer than $\mu+\sigma F^{-1}(\tau)$. Mathematically, the unreliability area is the summation of all unexpected delays (i.e., actual delays $\sigma F^{-1}(x)$ in excess of the expected delay $\sigma F^{-1}(\tau)$):

$$S_u = \sigma \int_\tau^1 \left( F^{-1}(x) - F^{-1}(\tau) \right) dx \tag{6}$$

---

[2] Such a partition between reliability and unreliability is more naturally applicable to flexible services but can also be applied to public transport services.

As shown in Eq. (6) and Figure 2, the unreliability area is mainly determined by the upper tail of the travel time percentile function. Highly-skewed travel time distributions with long/fat tails have been verified in many empirical studies (van Lint and van Zuylen, 2005; FHWA, 2006; van Lint *et al.*, 2008; Fosgerau and Fukuda, 2012; Susilawati *et al.*, 2013; Delhome *et al.*, 2015; Kim and Mahmassani, 2015; Zang *et al.*, 2018b; Li, 2019), thus highlighting the potential importance of unreliable aspect of travel time. To accommodate this unreliability in the arrival time, a risk-averse traveler must depart earlier than the original "optimal" departure time shown in Eq. (5), and this behavior has been verified by Xiao and Fukuda (2015) and Siu and Lo (2014). Using Stated Preference data, these authors found that risk-averse travelers are mostly pessimistic and tend to choose an earlier than optimal departure time. Whilst we cannot of course confirm as such, the fact that travelers' chose in this manner could suggest that travelers' base their preferences on both reliable and unreliable aspects of travel time[3]. This exactly corresponds to what our paper represents: travelers maximize expected utility in the context of both reliable and unreliable aspects of travel time.

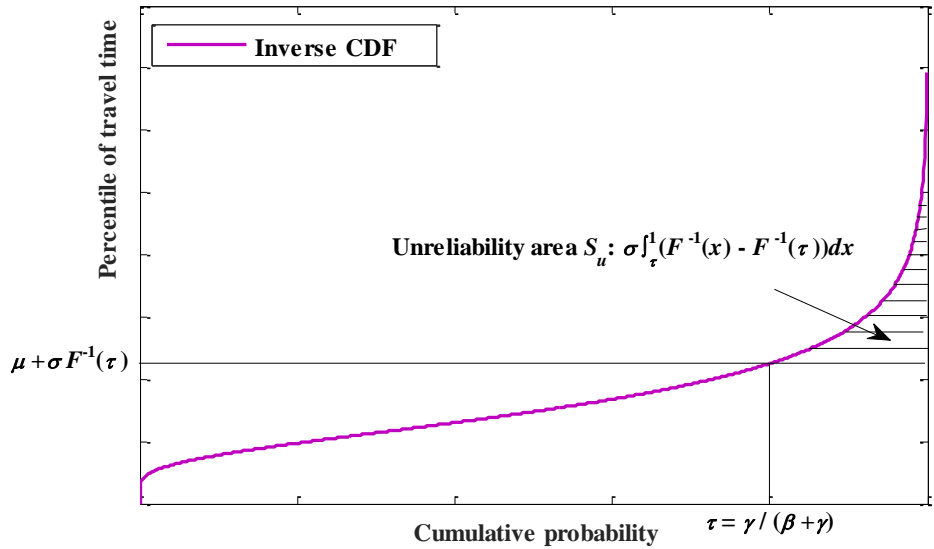

Figure 2. The inverse cumulative distribution function (CDF, i.e., percentile

---

[3] It is worthwhile to note that earlier departure time may also benefit users as they can attend some pre-event activities before their events (Bao et al., 2023).

function) of travel time and unreliability area.

Note that the present paper focuses only on risk-averse travelers unless otherwise specified, as the literature widely demonstrates the attitude of risk aversion toward travel time variability (e.g., Knight, 1974; Senna, 1994; Rietveld *et al*., 2001; Lo *et al*., 2006; Shao *et al*., 2006; Batley, 2007; Chen and Zhou, 2010; Li *et al*., 2010; de Palma *et al*., 2012; Sikka and Hanley, 2013; Zhang and Homem-de-Mello, 2017; Alonso-González *et al*., 2020, Li *et al*., 2022), and indeed the schedule delay model gives a linear-piecewise approximation to the concave utility function synonymous with risk aversion to travel time.

In subsequent sections, we show that the unreliability area is a key element of two well-defined measures accounting for unreliable aspects of travel time, namely the reliability premium (Batley, 2007) and the mean-excess travel time (Chen and Zhou, 2010). Specifically, we find that the "weight" attached to the unreliability area is the only difference between these two measures.

**2.2 Two Measures to Account for Travel Time Distribution Tail**

*(1) Reliability premium*

Adapting the concept of the risk premium (Pratt, 1964), Batley (2007) proposed the reliability premium as a measure of the cost of eliminating travel time variability (including both reliable and unreliable aspects of travel time) for a given departure time. Specifically, Batley (2007) derived the reliability premium for accepting a delayed but certain arrival time based on the concept of the certainty equivalent. Definition 1 below formally defines the concept of the certainty equivalent.

**Definition 1**. For the arrival time of a trip under uncertainty for a given departure time $D$ with expected utility $EU(D, A)$, its certainty equivalent, denoted by $A_C$, is the arrival time of a trip under certainty with the same departure time that satisfies $EU(D, A) = U(D, A_C)$.

Based on **Definition 1**, Batley (2007) defined the reliability premium as: *the 'reliability premium' measures, for a given departure time, the delay in arrival time that the individual would be willing-to-pay in exchange for eliminating the variability. The reliability premium thus measures the costs borne by the traveler that arise specifically from variability in arrival time.*

Accordingly, the reliability premium $\pi$ defined by Batley (2007) can be calculated by $EU(D_m, A) = U(D_m, A_C)$ where $A_C = A + \pi$. Nevertheless, **Definition 1** restricts the certainty equivalent to delayed arrival time for a given departure time of a trip under certainty. However, when facing travel time variability, in addition to adjusting arrival time, an individual can also adjust departure time. The present paper focuses on such a scenario, and **Definition 2** modifies the definitions of the certainty equivalent and reliability premium accordingly. Zang et al. (2022a) considered the above two scenarios as well as a more general scenario where both departure and arrival times are adjusted to eliminate the variability, which then defines the certainty equivalent and reliability premium more formally and fully on this basis than the present paper and Batley (2007).

**Definition 2**. Certainty equivalent and reliability premium for a given arrival time.
(i) For a trip under uncertainty with expected utility $EU$ for a given arrived time $A$, the certainty equivalent, denoted by $D_C$, is the outcome of a trip under certainty with utility $U$ that satisfies the following condition: $EU(D, A) = U(D_C, A)$.
(ii) To eliminate the travel time variability resulting from a given event $(D, A)$, the reliability premium $\pi$ is defined as the amount of early adjustment in the departure time $D$ compared to the expected departure time that a traveler is willing to pay to eliminate the travel time variability.

With Definition 2, the following Proposition 1 formally presents the statement of the reliability premium for the scenario of early adjustment of departure time.

**Proposition 1**. To eliminate the travel time variability resulting from a given arrival time $A$, the reliability premium $\pi$ is defined as the maximum amount of early adjustment to $D$, and its formulation is expressed as

$$\pi = \frac{\beta + \gamma}{\alpha} \left( \int_{\frac{-D-\mu}{\sigma}}^{\infty} (\mu + \sigma x + D) \cdot f(x) dx \right) \qquad (7)$$

Note that $(\beta + \gamma)$ also appears in the expected utility function (4). The $(\beta + \gamma)$ term is the sum of the marginal utilities of SDE and SDL, or alternatively the difference between the slopes of the linear piece-wise sections of the utility function before and after the PAT (Batley, 2007), giving rise to the slope of the corresponding expected utility function across the range of travel times. The term in the bracket on the right-hand-side of Eq. (7) is the integral of $T-D$ (i.e., lateness) over the travel time distribution. $\alpha$ is the marginal utility of travel time, which serves to report the reliability premium in travel time units.

*(2) Mean-excess travel time*

It can be shown that the optimal departure time Eq. (5) is exactly the travel time budget (TTB) defined in Lo *et al*. (2006) when $\tau = \gamma/(\beta + \gamma)$, i.e., pertaining to the reliability area in Figure 2. For a travel time distribution, the general definition of the TTB $b(\tau)$ can be given using the following chance-constraint model:

$$b(\tau) = \min\left\{\overline{T} \mid \Pr(T \leq \overline{T}) \geq \tau\right\} = \min\left\{\overline{T} \mid \Pr\left(X \leq \frac{\overline{T} - \mu}{\sigma}\right) \geq \tau\right\} \tag{8}$$

where $T$ is the travel time, $X$ is the standardized travel time, and $\overline{T}$ is the minimum travel time required to ensure the desired punctuality requirements $\tau$ (i.e. to satisfy the traveler's required probability of being on time).

For the continuous travel time distribution, the TTB is the $\tau$-percentile of travel time:

$$F\left(\frac{\overline{T}_{\min} - \mu}{\sigma}\right) = \tau \Rightarrow \overline{T}_{\min} = \mu + \sigma F^{-1}(\tau) \Rightarrow b(\tau) = \mu + \sigma F^{-1}(\tau) = \mu + \delta_{TTM} \tag{9}$$

where $\sigma F^{-1}(\tau)$ is the travel time margin $\delta_{TTM}$ corresponding to $\tau$, i.e., $\delta_{TTM} = \sigma F^{-1}(\tau)$.

From Eq. (9), we can see that the safety margin $\delta$, which is added by travelers to hedge against uncertainty in travel time, is further specified as the travel time margin $\delta_{TTM}$ in the TTB model. Thus, we have TTB = mean travel time + travel time margin. However, as already noted in Section 2.1 and shown in Figure 2, the travel time margin (i.e., $\delta_{TTM}$) in TTB considers only the reliable aspect of travel time variability, but does not reflect the unreliable aspect associated with trip times exceeding $b(\tau)$. To explicitly consider both the reliable and unreliable aspects of travel time, Chen and Zhou (2010) proposed the mean-excess travel time (METT), which can be defined as the summation of the expected excess delay (EED) and TTB, i.e., METT = TTB + EED. Note that the METT is an adaptation of the conditional value-at-risk (CVaR) concept from finance (Rockafellar and Uryasev, 2000, 2002) to the context of travel time variability. The mathematical expression of the EED, denoted by $\delta_{EED}$, is

$$\delta_{EED} = E\left[T - b(\tau) \mid T \geq b(\tau)\right] = \sigma \cdot \frac{\int_{\tau}^{1}\left(F^{-1}(x) - F^{-1}(\tau)\right)dx}{1 - \tau} = \sigma \cdot \zeta_{EED} \tag{10}$$

where $\delta_{EED}$ is the product of the standard deviation $\sigma$ and the EED scaling factor $\zeta_{EED}$.

**Proposition 2.** The relationship between the expected excess delay $\delta_{EED}$ in the METT and the unreliability area $S_u$ can be expressed as: $\delta_{EED} = S_u/(1 - \tau)$.

**Proof**. One can easily find this relationship according to Eq. (6) and Eq. (10).

Proposition 2 shows that, compared to the TTB, the METT considers the unreliable aspect of travel time by adding the EED, i.e., attaching a weight parameter to the unreliability area. Furthermore, the EED is actually the *mean value* of all unexpected delays and intuitively addresses the question of "how bad should I *expect* the unreliable aspect of travel time to be?". As $b(\tau) = \mu + \delta_{TTM}$, the METT can be decomposed into three parts: mean travel time, travel time margin and EED, and can be rewritten as

$$\eta(\tau) = b(\tau) + \delta_{EED} = \mu + \delta_{TTM} + \delta_{EED} = \mu + \sigma\zeta_{ETT} \qquad (11)$$

where $\sigma\zeta_{ETT}$ is the excess travel time (ETT) in the METT and is calculated by summing the travel time margin and EED, i.e., $\sigma\zeta_{ETT} = \delta_{TTM} + \delta_{EED}$. $\zeta_{ETT} = \dfrac{\beta+\gamma}{\beta} \int_{\frac{\gamma}{\beta+\gamma}}^{1} F^{-1}(x)\,dx$.

$\zeta_{ETT}$ is interpreted as the ETT scaling factor because ETT is the product of $\sigma$ and $\zeta_{ETT}$.

Based on Eq. (11), Figure 3 illustrates how both the reliable and unreliable aspects of travel time are incorporated into the METT. Figure 3 clearly shows that the METT considers both reliable aspect through the travel time margin $\delta_{TTM}$ and unreliable aspect through the EED $\delta_{EED}$. That said, the above decomposition of the METT is used simply for exposition, and it does not mean that the METT is determined segmentally or as a threshold-based measure. Instead, the METT is defined as a conditional expectation beyond the travel time budget, which is expressed as:

$$\eta(\tau) = \frac{1}{1-\tau}\int_{\tau}^{1} b(x)\,dx = \mu + \frac{1}{1-\tau}\sigma \cdot \int_{\tau}^{1} F^{-1}(x)\,dx \qquad (12)$$

It can thus be seen from Eq. (12) that the METT is a function of the probability $\tau$.

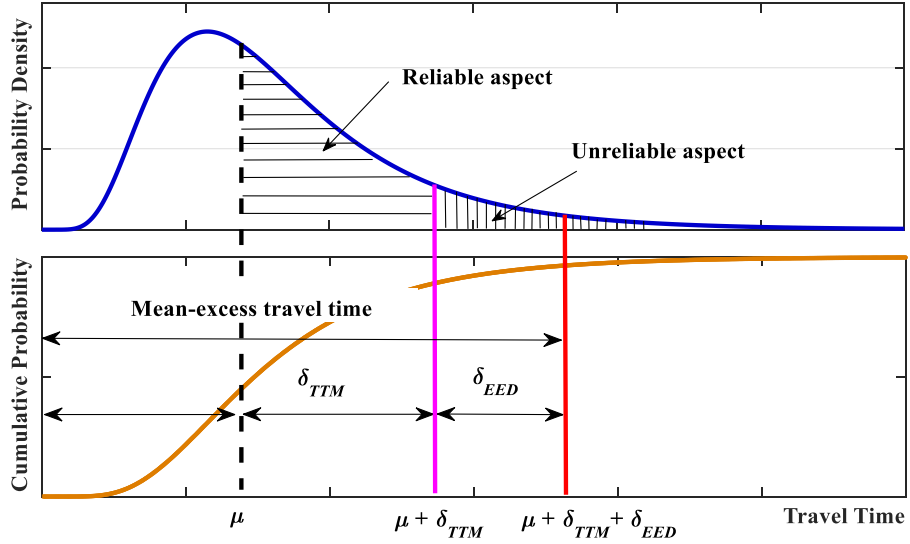

Figure 3. The illustration of reliable and unreliable aspects considered by the METT.

Proposition 3 below shows a proportional relationship between the expected excess delay and reliability premium if the departure time is the optimal departure time, i.e., $D = D^* = -b(\tau)$ when $\tau = \gamma/(\beta + \gamma)$.

**Proposition 3**. The expected excess delay $\delta_{EED}$ is $\alpha/\beta$ times the reliability premium $\pi$ if $D = -b(\tau)$. Namely,

$$\delta_{EED} = \frac{\alpha}{\beta}\pi \tag{13}$$

An examination of Eqs. (B-3) and (B-4) reveals that the expressions for the EED and reliability premium differ only in terms of the "weight" assigned to the unreliability area $S_u$. Therefore, we can establish a proportional relationship between the reliability premium and the EED as given by Proposition 3 and then use the EED to represent the reliability premium and vice versa. Given this proportional relationship between the EED and the reliability premium, the remainder of the paper will focus upon the implementation of EED within METT to quantify the distribution tail cost of travel time. Note that the METT has been widely used in the literature, in contexts as diverse as stochastic perception error (Chen *et al*., 2011; Xu *et al*., 2013), hazardous material routing on time-dependent networks (Toumazis and Kwon, 2013, 2016; Su and Kwon, 2020), travel time robust reliability (Sun and Gao, 2012), strategy costs in schedule-based transit networks (Rochau *et al*., 2012), risk-based transit schedule design (Zhao *et al*., 2013), network performance assessment (Xu *et al*., 2014, 2021), toll pricing (Feyzioğlu and Noyan, 2017), and speed limit evaluation (Xu *et al*., 2018).

## 2.3 Value of Travel Time Distribution Tail

As just discussed, under the METT criterion, a traveler would consider both reliable and unreliable factors when choosing a departure time, and thus with reference to Eqs. (5), (9) and (11), the departure time is given as

$$D_{METT} = -\mu - \delta_{TTM} - \delta_{EED}$$
$$= -\mu - \sigma F^{-1}\left(\frac{\gamma}{\beta+\gamma}\right) - \sigma \cdot \frac{\beta+\gamma}{\beta} \int_{\frac{\gamma}{\beta+\gamma}}^{1}\left(F^{-1}(x) - F^{-1}\left(\frac{\gamma}{\beta+\gamma}\right)\right)dx \quad (14)$$

Then, based on Eq. (4), the expected utility of a trip under the METT criterion is

$$EU(D_{METT}, T) = -\left\{\alpha\mu + \beta\sigma\zeta_{ETT} + (\beta+\gamma)\left(\int_{\zeta_{ETT}}^{\infty}(\mu + \sigma x - \mu - \sigma\zeta_{ETT}) \cdot f(x)dx\right)\right\}$$
$$= -\left\{\alpha\mu + \beta\sigma\zeta_{ETT} + (\beta+\gamma)\sigma\left(\int_{\zeta_{ETT}}^{\infty}(x - \zeta_{ETT}) \cdot f(x)dx\right)\right\} \quad (15)$$
$$= -\left\{\alpha\mu + \beta\sigma\zeta_{ETT} + (\beta+\gamma)\sigma\left(\int_{F(\zeta_{ETT})}^{1}F^{-1}(x)dx - \zeta_{ETT} + \zeta_{ETT}F(\zeta_{ETT})\right)\right\}$$

where $\zeta_{ETT} = \frac{\beta+\gamma}{\beta}\int_{\frac{\gamma}{\beta+\gamma}}^{1}F^{-1}(x)dx$.

Compared to the optimal departure time $D^*$ given by Eq. (5) (abbreviated as the TTB criterion since $D^* = -b(\tau)$) that considers only the reliable aspect of travel time, the EED under the METT criterion represents the amount by which departure time must be brought forward in order to fully mitigate for unreliable aspect of travel time. Intuitively, the rescheduling of the departure time imposes a cost on the traveler, since it interrupts whatever activity was previously conducted at that time. Proposition 4 uses the relationship between the expected trip utility under the METT criterion and the expected trip utility under the TTB criterion to assist in quantifying this rescheduling cost.

**Proposition 4.** For trip utility under the METT criterion and TTB criterion, we have

$$|EU(D_{METT}, T)| \geq |EU(D_{TTB}, T)| \quad (16)$$

Proposition 4 suggests that the trip cost under the METT criterion is always greater than or equal to that under the TTB criterion. Therefore, the difference, $|EU(D_{METT}, T)| - |EU(D_{TTB}, T)|$, is the cost the traveler must pay to account for the distribution tail (i.e., unreliable aspect) of travel time when choosing the departure time. As the EED is used to quantify the distribution tail of travel time, the value of travel time distribution tail (VODT) can be defined as the ratio of the rescheduling cost to the EED thus

$$VODT = \frac{|EU(D_{METT}, T)| - |EU(D_{TTB}, T)|}{|D_{METT} - D_{TTB}|} = \frac{|EU(D_{METT}, T)| - |EU(D_{TTB}, T)|}{\delta_{EED}}$$

$$= (\beta + \gamma) \frac{\zeta_{ETT} F(\zeta_{ETT}) - \zeta_{ETT} + \int_{F(\zeta_{ETT})}^{1} F^{-1}(x) dx}{\zeta_{EED}} \quad (17)$$

In other words, the cost of travel time distribution tail is the product of the VODT and the EED. The non-negativity of the VODT is guaranteed by Proposition 4, and therefore we have the condition $l > 1$, where $l = \int_{F(\zeta_{ETT})}^{1} F^{-1}(x) dx \Big/ \zeta_{ETT}(1 - F(\zeta_{ETT}))$. Similarly, as $\delta_{TTM}$ is used to characterize the reliable aspect of travel time, we can obtain what we define as the value of travel time reliability (VOR) in the Introduction in terms of per unit of $\delta_{TTM}$ (rather than variance or standard deviation) as

$$VOR = \frac{|EU(D_{TTB}, T)| - |U(-\mu, \mu)|}{|D_{TTB} - (-\mu)|} = \frac{|EU(D_{TTB}, T)| - |U(-\mu, \mu)|}{\delta_{TTM}}$$

$$= \frac{\beta + \gamma}{F^{-1}(\gamma/(\beta + \gamma))} \int_{\frac{\gamma}{\beta + \gamma}}^{1} F^{-1}(p) dp \quad (18)$$

where $U(-\mu, \mu) = -\alpha\mu$ is the utility at the certainty.

**Remark 1.** Changing $\gamma$ from a constant to be an increasing function with severity of delay is an alternative way to quantify the VODT, but a balance needs to be established between the relative costs of expected delay and unexpected delay – which does not seem straightforward. Even if $\gamma$ is constant, the total penalty for late delay $\gamma SDL$ still increases with the severity of delay, which forces the traveler to depart earlier. Moreover, adjusting departure time via the METT criterion is very simple to take account of both expected and unexpected delays. Thus, this paper uses a constant $\gamma$ for simplicity.

By rearranging the terms in Eq. (15), we obtain

$$EU(D_{METT}, T) = -\left\{ \alpha\mu + \sigma\zeta_{ETT} \left( -\gamma + (\beta + \gamma) F(\zeta_{ETT}) + (\beta + \gamma) \frac{\int_{F(\zeta_{ETT})}^{1} F^{-1}(x) dx}{\zeta_{ETT}} \right) \right\} \quad (19)$$

Recall that, in contrast to the travel time margin within TTB which captures the reliable aspect of travel time variability, the ETT within METT combines the travel time margin and the unreliability measure (i.e., EED) to capture the unreliable aspect of travel time variability. Hence, the value of travel time variability (VOV) in terms of per unit of ETT can be defined as

$$VOV = \frac{|EU(D_{METT}, T)| - |U(-\mu, \mu)|}{|D_{METT} - (-\mu)|} = \frac{|EU(D_{METT}, T)| - |U(-\mu, \mu)|}{\sigma \zeta_{ETT}} \quad (20)$$

$$= -\gamma + (\beta + \gamma) F(\zeta_{ETT}) + \frac{\beta + \gamma}{\zeta_{ETT}} \int_{F(\zeta_{ETT})}^{1} F^{-1}(x) dx$$

By comparing Eqs. (17), (18) and (20), it is easy to see that the VOV arises from the weighted average of the VOR and VODT, i.e., $VOV = (VOR \cdot \delta_{TTM} + VODT \cdot \delta_{EED})/(\delta_{TTM} + \delta_{EED})$, thereby capturing the value of travel time variability in terms of both reliable and unreliable aspects.

Proposition 5 below investigates the impact of the preference parameters on the VOV, and its corollary theoretically proves that the VOV exhibits a diminishing marginal effect under a validity condition. Before presenting Proposition 5, Lemma 1 gives the preliminaries used to derive this proposition.

**Lemma 1.** The ETT scaling factor is a strictly non-increasing function of $\beta$ and a strictly non-decreasing function of $\gamma$, namely

$$\frac{\partial \zeta_{ETT}}{\partial \beta} \leq 0, \text{ and } \frac{\partial \zeta_{ETT}}{\partial \gamma} \geq 0 \quad (21)$$

**Proposition 5.** For the VOV in terms of per unit of ETT, we have
(1) $\frac{\partial VOV}{\partial \beta} \geq 0$; and (2) $\frac{\partial VOV}{\partial \gamma} \leq 0$ if $l \leq \kappa + 1$, where $\kappa = \zeta_{EED}/F^{-1}(\tau)$.

Recall that the probability of arriving on-time is given by $\tau = \gamma/(\beta + \gamma)$. Generally speaking, travelers increase $\tau$ by placing a larger penalty on the SDL, i.e., increasing $\gamma$. Under this assumption, Corollary 1 can be concluded according to Proposition 5.

**Corollary 1.** The VOV is a strictly non-increasing function of the punctuality requirement, namely $\frac{\partial VOV}{\partial \tau} \leq 0$ if $l \leq \kappa + 1$.

**Proof.** $\frac{\partial \tau}{\partial \gamma} = \frac{\beta}{(\beta + \gamma)^2} > 0$. If $l \leq \kappa + 1$, then

$$\frac{\partial VOV}{\partial \tau} = \frac{\partial VOV}{\partial \gamma} \cdot \frac{1}{\frac{\partial \tau}{\partial \gamma}} \leq 0 \quad (22)$$

This completes the proof. □

**Remark 2**. As $l \leq \kappa + 1$ (hereinafter we will refer to this as the validity condition) is required in Proposition 5 and Collary 1, it is important to investigate the meaning of $l$ and $\kappa + 1$ and the feasibility of the condition $l \leq \kappa + 1$. By rewriting the formulations of $l$ and $\kappa + 1$, we have:

$$\kappa + 1 = \frac{\zeta_{EED} + F^{-1}(\tau)}{F^{-1}(\tau)} = \frac{\zeta_{ETT}}{\zeta_{TTM}}$$

$$l = \frac{\sigma \int_{F(\zeta_{ETT})}^{1} F^{-1}(x) dx}{1 - F(\zeta_{ETT})} \frac{1}{\frac{\sigma \int_{\tau}^{1} F^{-1}(x) dx}{1 - \tau}} = \frac{E[X > \zeta_{ETT}]}{E[X > \zeta_{TTM}]} \quad (23)$$

Therefore, $\kappa + 1$ is actually the ratio of the ETT scaling factor $\zeta_{ETT}$ to the TTM scaling factor $\zeta_{TTM}$, while $l$ is the ratio of the mean value of all standardized travel time $X$ beyond $\zeta_{ETT}$ to the mean value of all $X$ beyond $\zeta_{TTM}$. For the feasibility of $l \leq \kappa + 1$, Appendix D theoretically proves that $l \leq \kappa + 1$ is true if $F^{-1}(x)'' / F^{-1}(x)' \leq \frac{1}{1-x}$ for $x \in [\tau, 1)$. Specifically, if $F^{-1}(x)$ is a concave function, $l \leq \kappa + 1$ must be true. □

The numerical examples in Section 3 of this paper demonstrate that $l \leq \kappa + 1$ is a relatively slack condition that can be easily satisfied by realistic travel time datasets. Therefore, Proposition 5 and the corresponding Corollary 1 indicate that with an increase in $\gamma$, the ETT added by a traveler to hedge against travel time variability increases while the VOV in terms of per unit of ETT decreases. This indicates a diminishing marginal return of an increase in the ETT. In other words, for risk-averse travelers, the VOV has a *diminishing marginal effect*. In a broader picture, the diminishing marginal effect has also been observed in other related scenarios. For instance, in terms of the value of travel time saving, Metz (2008) found that the additional benefit from further travel time savings tend to decline, which is a case of diminishing marginal effect in terms of travel time savings. In addition, for the reliable network design problem, both Chen *et al*. (2007) and Xu *et al*. (2014) found a diminishing marginal effect of network reliability performance improvement in terms of construction budget. The next section proposes the travel time variability ratio based on the diminishing marginal effect of VOV to provide insights for travelers in determining a reasonable punctuality requirement $\tau$.

### 2.4 Travel Time Variability Ratio

Travelers may well understand that an increase in their punctuality requirements $\tau$ means that they are more risk averse towards travel time variability and must depart earlier. Nonetheless, it is difficult for travelers to explicitly know the cost of any such increase in *τ*. Using the value of travel time (VOT) as a reference, we propose the travel time variability ratio based on the VOV as a means of representing this cost.

The travel time variability ratio (TTVR) is defined as the ratio of the VOV to the VOT,

$$\rho_{TTVR} = \frac{VOV}{VOT} = -\frac{\gamma}{\alpha} + \frac{\beta+\gamma}{\alpha} F(\zeta_{ETT}) + \frac{\beta+\gamma}{\alpha} \frac{\int_{F(\zeta_{ETT})}^{1} F^{-1}(x)dx}{\zeta_{ETT}} \quad (24)$$

This definition is similar to that of the travel time reliability ratio (TTRR), firstly proposed by Jackson and Jucker (1981), and later elaborated upon by Fosgerau (2017) on the basis of important conceptual and theoretical developments of the VOR (Fosgerau and Karlström, 2010). As noted earlier, the travel time reliability ratio is defined as the ratio of the VOR (in terms of per unit standard deviation) to the VOT, namely $\rho_{TTRR} = \frac{\beta+\gamma}{\alpha} \int_{\frac{\gamma}{\beta+\gamma}}^{1} F^{-1}(x)dx$. The travel time reliability ratio is a typical valuation method used to include the VOR in route choice models and transportation scheme appraisal (Fosgerau, 2017; OECD, 2016; Taylor, 2017).

Here, we make some observations on the travel time variability ratio. First, this dimensionless measure is intended to increase the intuitive comprehensibility of the VOV, through reference to the VOT. Second, as in Eq. (24), the computational challenges of calculating the travel time variability ratio are the same as those of calculating the travel time reliability ratio (Zang *et al.*, 2018a), i.e., the complex integral term and the inverse CDF. The travel time variability ratio can be calculated efficiently and effectively, based on the analytical estimation method developed by Zang *et al.* (2018b) for TTRR. Finally, recall that *l* > 1 and the lower bound of the travel time variability ratio is

$$\rho_{TTVR} \geq -\frac{\gamma}{\alpha} + \frac{\beta+\gamma}{\alpha} F(\zeta_{ETT}) + \frac{\beta+\gamma}{\alpha}(1 - F(\zeta_{ETT})) = \frac{\beta}{\alpha} \quad (25)$$

Given this lower bound, the VOV is always greater than the value of the SDE, consistent with the assumption that *γ* > *β* for risk-averse travelers.

Proposition 6 below explores the influence of the preference parameters on the travel time reliability ratio and the travel time variability ratio for risk-averse travelers.

**Proposition 6.** For the travel time reliability ratio and travel time variability ratio,

(1) $\frac{\partial \rho_{TTRR}}{\partial \alpha} < 0$ and $\frac{\partial \rho_{TTVR}}{\partial \alpha} < 0$;

(2) $\frac{\partial \rho_{TTRR}}{\partial \beta} \geq 0$ and $\frac{\partial \rho_{TTVR}}{\partial \beta} \geq 0$;

(3) $\frac{\partial \rho_{TTRR}}{\partial \gamma} \geq 0$ and $\frac{\partial \rho_{TTVR}}{\partial \gamma} \leq 0$ if $l \leq \kappa + 1$.

**Proof.** Li (2019) has provided a proof of the effects of the preference parameters on the travel time reliability ratio. The proof of these effects on the travel time variability ratio is mainly based on the proof of Proposition 5. Please refer to the proof in Appendix B. This completes the proof. □

As demonstrated by Proposition 6, $\alpha$ and $\beta$ (i.e., the marginal utilities of travel time and SDE, respectively) affect the travel time reliability ratio and travel time variability ratio similarly, whereas $\gamma$ (i.e. the marginal utility of SDL) affects these ratios differently. Specifically, the travel time reliability ratio is positively related to $\gamma$, whereas the travel time variability ratio is negatively related to $\gamma$ under $l \leq \kappa + 1$. In other words, as the traveler increases excess travel time (ETT) to mitigate for unreliability aspect and thereby increase the probability of arriving on-time, the value of further reducing travel time variability diminishes. On this basis, the proposed travel time variability ratio can help travelers to understand the cost of improving their punctuality requirements and thereby achieve a better balance between on-time arrival and trip cost, to be illustrated in detail by the numerical example in Section 3.2.

## 3. NUMERICAL EXAMPLES

In this section, we use an illustrative network to demonstrate the importance of considering the cost of travel time distribution tail. Then, we use several datasets involving different road types and traffic states to test the degree of slackness of the validity condition in Propositions 5 and 6. Finally, we illustrate the application of the travel time variability ratio.

### 3.1 Illustrative Example of Including the Cost of Travel Time Distribution Tail

Note that this illustrative example is not used to prove or claim that our proposed method is better than the existing methods in the literature. Rather we use this example to highlight the importance of including the cost of distribution tail within the cost of travel time variability, because of the long distribution tail of travel time.

As shown in Figure 4, a hypothetical network in which a single origin–destination pair is connected by six parallel paths is used for this illustration. In this example, the punctuality requirement is assumed to be 80%, so as to be consistent with Small (1982), Fosgerau and Karlström (2010) and Taylor (2017), i.e., the probability of arriving on or before the PAT is given by $\tau = 0.8$ and $\gamma / (\beta + \gamma) = 0.8$; but there is no practical constraint on what value $\tau$ should take. We also assume that the random travel times of the six paths follow the lognormal distribution $Logn(\xi, \psi)$, which enables a good characterization of a right-skewed travel time distribution (Emam and Al-Deek, 2006; Kaparias *et al*., 2008; Rakha *et al*., 2010; Arezoumandi, 2011; Chen *et al*., 2014). For each route *p*, the lognormal distribution parameters $\xi_p$ and $\psi_p$ are given in Figure 4, and the PDF is visualized in Figure 5. From Path 1 to Path 6, the mean travel time decreases, and the skewness of the travel time distribution and the length of the distribution tail increase, thereby illustrating the effect of a longer or fatter distribution tail on the cost of distribution tail.

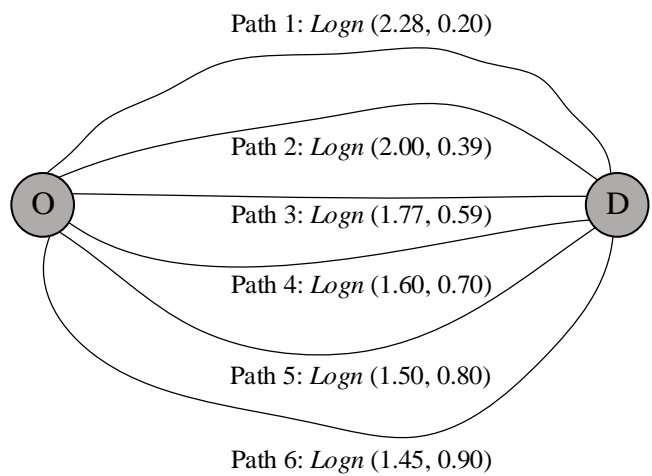

Figure 4. Hypothetical network with one origin–destination pair, connected by six paths.

To explore the impact of including reliable and unreliable aspects of travel time within the trip cost, the following three scenarios are considered.

- **Scenario 1**: The traveler only wants to minimize their trip cost in terms of the mean travel time, and his/her trip cost includes only the cost of the certainty equivalent (hereinafter we refer to this more succinctly as the certainty cost).
- **Scenario 2**: The traveler pays attention to reliable aspect of travel time and minimizes the trip cost in terms of TTB, and his/her trip cost includes the certainty cost and reliability cost.
- **Scenario 3**: The traveler pays attention to both reliable and unreliable aspects of travel time and minimizes the trip cost in terms of METT, and his/her trip cost includes the certainty cost, reliability cost and tail cost.

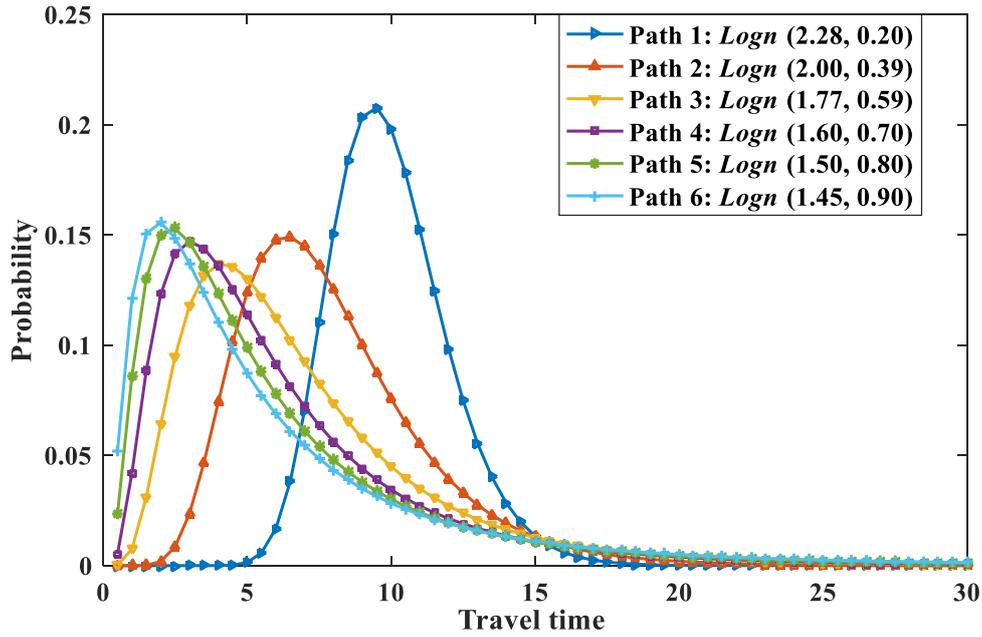

Figure 5. Probability distribution functions of the six lognormal distributions.

Table 1 gives the mathematical formulations used to calculate the three variants of trip costs, and Table 2 presents the derived trip costs of the six paths under the three scenarios. In the "Trip cost [TTB]" column of Table 2, trip cost is expressed as the summation of the certainty cost and reliability cost. Similarly, in the "Trip cost [METT]" column, trip cost is expressed as the summation of the certainty cost, reliability cost, and tail cost.

Table 1. Mathematical formulations of trip costs under different scenarios.

| Scenario | Mathematical formulation |
| --- | --- |
| 1: Trip cost [Mean] | $c_p = \alpha \mu$ |
| 2: Trip cost [TTB] | $c_p = \alpha \mu + \delta_{TTM} \cdot VOR$ |
| 3: Trip cost [METT] | $c_p = \alpha \mu + \delta_{TTM} \cdot VOR + \delta_{EED} \cdot VOU$ |

Clearly, different considerations of the trip cost would yield different selections of the best and worst routes. For example, under Scenario 1, Path 5 would be the best choice for travelers, with the lowest cost of 12.36 as shown in Table 2. However, if travelers paid attention to travel time reliability, the trip cost [calculated in terms of TTB] would include both the certainty cost and reliability cost. Accordingly, a reliability cost of 7.01 would need to be added to Path 5, and the resulting trip cost [TTB] would be greater than that of Path 4. Therefore, Path 4 would be preferable under Scenario 2. Under Scenario 3, although Path 4 has the minimum trip cost [METT], the tail cost of this path (i.e., 1.60) is much larger than that of Path 1 (i.e., 0.39) or Path 2 (i.e., 0.80). As for Path 1 and Path 6, Path 1 yields the highest trip cost [Mean], while the certainty cost of Path 6 is only 0.64 times (i.e., 12.80/19.94) that of Path 1. However, when including both the reliability cost and tail cost, Path 6 yields the highest trip cost [METT].

Table 2. Coefficients of variation (CoV) and trip costs [calculated by the formulae in Table 1] under three scenarios for six paths.

| Path ID | CoV | Scenario 1: Trip cost [Mean] | Scenario 2: Trip cost [TTB] | Scenario 3: Trip cost [METT] |
|---|---|---|---|---|
| 1 | 0.20 | 19.94 | 22.35 (19.94+2.41) | 22.74 (19.94+2.41+0.39) |
| 2 | 0.41 | 15.93 | 19.94 (15.93+4.01) | 20.74 (15.93+4.01+0.80) |
| 3 | 0.64 | 13.93 | 19.51 (13.93+5.58) | 20.84 (13.93+5.58+1.32) |
| 4 | 0.79 | 12.65 | **18.80** (12.65+6.15) | **20.41** (12.65+6.15+1.60) |
| 5 | 0.94 | **12.36** | 19.37 (12.36+7.01) | 21.35 (12.36+7.01+1.98) |
| 6 | 1.10 | 12.80 | 21.01 (12.80+8.21) | 23.53 (12.80+8.21+2.51) |

Note: CoV shows the extent of travel time variability in relation to mean travel time

To further explore the effects of including both the reliability cost and tail cost, the percent values of the certainty cost, reliability cost, and tail cost relative to the total trip cost for all six paths under Scenario 3 are presented in Figure 6. Recall that from Path 1 to Path 6, the mean travel time decreases, while the skewness and tail length of the travel time distribution increase. As shown in Figure 6, the percent of certainty cost decreases monotonically with decreasing mean travel time, whereas the percent of reliability cost and percent of tail cost increase monotonically with increasing travel time distribution skewness. Specifically, from Path 1 to Path 6, the percent of reliability cost increases from 11% to 35%, and the percent of tail cost increases from 2% to 11%.

In other words, for a more highly skewed travel time distribution with a longer tail, both the reliability cost and tail cost will play a more significant role in the total trip cost. We further verified this in 13 realistic datasets (Section 3.2), and the minimum and maximum percent of tail costs are 1.21% and 8.19%, respectively. Note that the literature documents extensive examples of right-skewed travel time distributions with long/fat tails, and the above findings indicate that ignoring the cost of travel time distribution tail may increase the risk of bias. Therefore, it is necessary to consider the cost of travel time distribution tail in modeling travelers' departure time and route choices, as well as in project appraisals. The concept and method of calculating the VODT proposed in this paper can support such requirements.

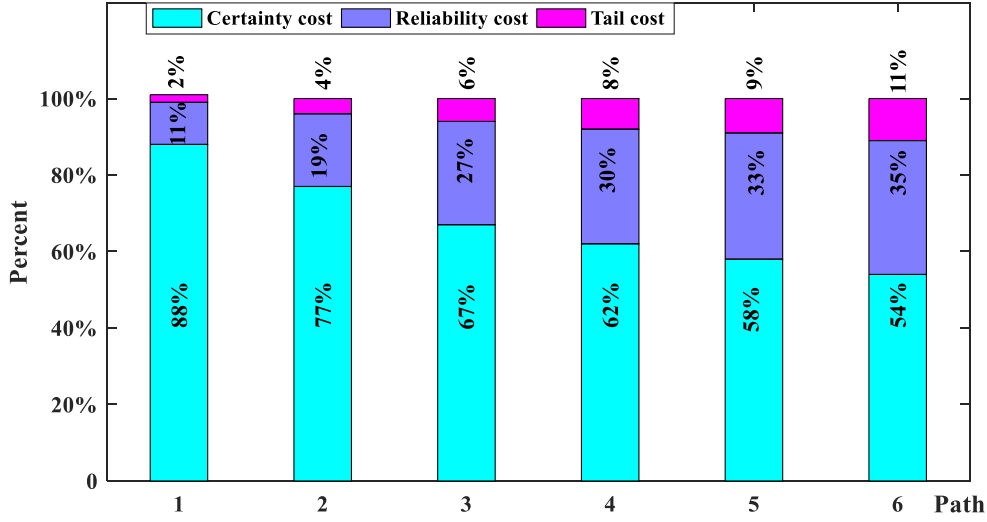

Figure 6. Percent values of the certainty cost, reliability cost, and tail cost relative to the total trip cost [METT] for six paths (Table 1 presents the formulae of calculating these costs).

## 3.2 The Feasibility of the Validity Condition in Propositions 5 and 6

Due to $\tau = \gamma/(\beta + \gamma)$, there is an unknown parameter $\tau$ in the validity condition $l \leq \kappa + 1$ in Propositions 5 and 6. Therefore, to explore the feasibility of the validity condition $l \leq \kappa + 1$ in practical applications, we use empirical datasets to examine whether $l \leq \kappa + 1$ is true for any $\tau$ between 0.5 and 1. Thirteen datasets involving different road types and traffic states were used. Eight datasets were obtained from the open Next Generation Simulation (NGSIM) dataset (NGSIM, 2005), and five were collected from the GPS records of the Tongji University School Bus. The thirteen datasets are described in Table 3 and the accompanying notes. We should point out that every record of route travel time in our data is directly obtained from the whole trajectory of one particular car or school bus along the whole route. This means that our route travel time datasets do not involve link segment combination, i.e., aggregating the travel time from the link level to the route level. For the methods for modelling route travel time distributions involving correlations between links, readers can refer to a recent review paper (Zang *et al*., 2022b).

Table 3. Detailed description of the 13 datasets.

| Dataset | Road type | Length (km) | Mean (min) | Standard deviation | Skewness | Kurtosis |
|---|---|---|---|---|---|---|
| Campus 1 | Expressways and urban streets | 37.10 | 59.90 | 7.76 | 1.09 | 3.09 |
| Campus 2 | | 37.10 | 57.80 | 7.16 | 1.47 | 2.70 |
| Campus 3 | | 37.10 | 51.97 | 7.56 | 2.44 | 8.57 |
| Campus 4 | | 37.10 | 65.71 | 11.13 | 1.21 | 2.78 |
| Campus 5 | | 37.10 | 80.79 | 17.09 | 0.66 | -0.49 |

| Dataset | Road type | Length (km) | Mean (s) | Standard deviation | Skewness | Kurtosis |
|---|---|---|---|---|---|---|
| Link 1 | Urban streets | 0.50 | 36.57 | 25.23 | 0.37 | -0.78 |
| Link 2 | | 0.50 | 15.24 | 9.93 | 2.80 | 10.86 |
| Link 3 | | 0.50 | 15.85 | 13.45 | 1.67 | 1.76 |
| Path | | 1.50 | 84.60 | 28.11 | 0.30 | 0.04 |

| Dataset | Road type | Length (km) | Mean (s) | Standard deviation | Skewness | Kurtosis |
|---|---|---|---|---|---|---|
| 101_1 | Freeway | 0.64 | 52.60 | 13.51 | 0.94 | 1.35 |
| 101_2 | | 0.64 | 68.42 | 21.79 | 0.40 | -0.89 |
| 101_3 | | 0.64 | 80.84 | 18.88 | -0.26 | 0.50 |
| 101_4 | | 0.64 | 65.23 | 21.38 | 0.52 | -0.52 |

a) School bus datasets. The GPS records of the school bus route from Jiading Campus to Siping Campus were collected between September 18, 2017 and January 17, 2018. The route covers both urban streets and expressways, and has an approximate length of 37.10 km. The school bus departs according to a timetable, and the five datasets (i.e., Campus 1 to Campus 5 in Table 3) used in this paper correspond to departures of 12:15, 14:00, 15:30, 16:30, and 17:20, respectively.

b) NGSIM Lankershim datasets. Northbound traffic datasets were collected at Lankershim Street on June 16, 2005. The studied road segment (path) consists of three links: Link 1 from Intersection No. 1 to No. 2, Link 2 from Intersection No. 2 to No. 3, and Link 3 from Intersection No. 3 to No. 4. These data yield three link-level datasets and one path-level dataset (i.e., Links 1 to 3 and Path in Table 3).

c) NGSIM Highway 101 datasets. The study site is a section of freeway with an approximate length of 640 meters. Southbound traffic data were collected between 07:50 and 08:05 on June 15, 2005. These data yield three datasets (i.e., 101_1 to 101_3 in Table 3) corresponding to 5-minute intervals and one dataset (i.e., 101_4 in Table 3) corresponding to the entirety of the 15-minute observation period.

An empirical dataset may not represent the upper tail very well, depending on the "extreme" travel time values recorded in the dataset. Rather, it may be better to fit the empirical dataset to obtain a more statistically representative distribution tail because the concept of the value of distribution tail depends on the tail of the distribution. Here, to determine whether the thirteen datasets satisfy the validity condition in Propositions 5 and 6, three representative methods, namely the lognormal distribution, Burr distribution (Susilawati *et al.*, 2013; Taylor, 2017), and analytical estimation method (Zang *et al.*, 2018b), were used to fit the empirical datasets. Taking the lognormal distribution as an example, the results show that the ratios of $\kappa+1$ to $l$ for the 13 datasets are all greater than or equal to 1, which is clearly shown by Figure 7. Therefore, all 13 datasets satisfy the validity condition $l \leq \kappa+1$ when using the lognormal distribution to fit empirical datasets.

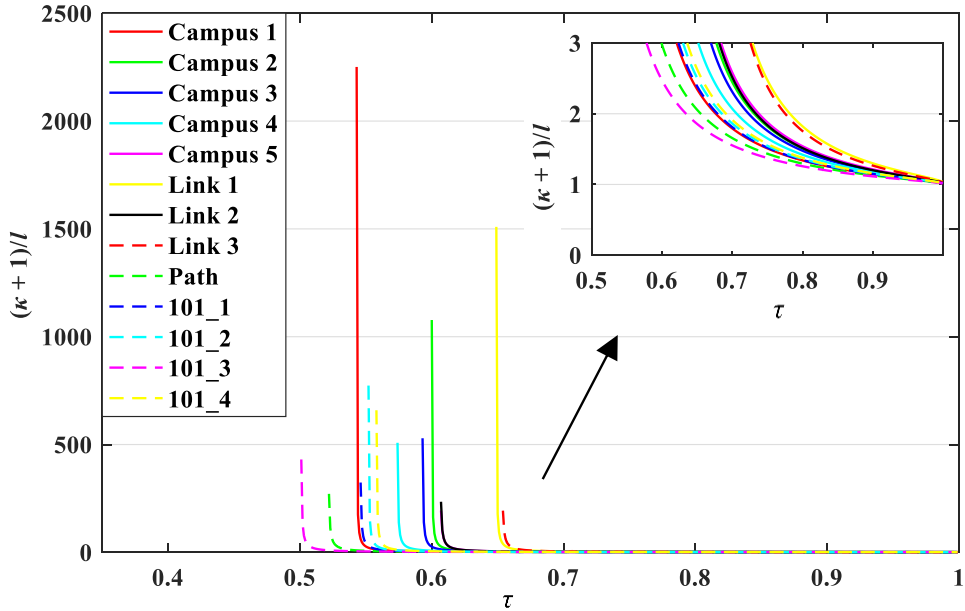

Figure 7. The values of the ratio of $\kappa+1$ to $l$ under different $\tau$ for 13 datasets.

In fact, we also reach the same conclusion when using the Burr distribution and the analytical estimation method to fit empirical datasets, and Table 4 summarizes the results. In other words, all empirical datasets satisfy the validity condition in Propositions 5 and 6. Note that the 13 datasets cover different road types and have different statistical characteristics including left-skewed, closely normal, and right-skewed distributions (van Lint *et al.*, 2008; Zang *et al.*, 2018b) as shown in Table 3. We can conclude that the validity condition is relatively slack and easily satisfied by realistic travel time datasets.

Table 4. Satisfaction of the validity condition $l \leq \kappa + 1$

| Dataset | Satisfy validity condition? | | | Dataset | Satisfy validity condition? | | |
|---|---|---|---|---|---|---|---|
| | Logn | Burr | Analytical | | Logn | Burr | Analytical |
| Campus_1 | ✓ | ✓ | ✓ | Link 3 | ✓ | ✓ | ✓ |
| Campus_2 | ✓ | ✓ | ✓ | Path | ✓ | ✓ | ✓ |
| Campus_3 | ✓ | ✓ | ✓ | 101_1 | ✓ | ✓ | ✓ |
| Campus_4 | ✓ | ✓ | ✓ | 101_2 | ✓ | ✓ | ✓ |
| Campus_5 | ✓ | ✓ | ✓ | 101_3 | ✓ | ✓ | ✓ |
| Link 1 | ✓ | ✓ | ✓ | 101_4 | ✓ | ✓ | ✓ |
| Link 2 | ✓ | ✓ | ✓ | | | | |

Note: ✓ indicates that the validity condition is satisfied.

### 3.3 Illustration and Application of the Travel Time Variability Ratio

This example illustrates how the travel time variability ratio empirically changes with the punctuality requirements $\tau$ and, more practically, how it can help travelers to achieve a better balance between trip reliability and trip cost. The corresponding theoretical basis has been given in Propositions 5 and 6 and Corollary 1.

We use the Highway 101_1 dataset as an example to calculate the travel time variability ratio based on the analytical estimation method (Zang *et al*., 2018a, 2018b). As for the parameters within the calculation, $\alpha$ can be viewed as a fixed value because (1) the VOT of any given risk-averse traveler for any given trip is usually constant, and (2) $\tau$ is independent of $\alpha$. Besides, to improve travel time reliability, a risk-averse traveler usually attaches a higher penalty to late arrival (i.e., a larger $\gamma$) so as to ensure tighter punctuality requirements $\tau$. Recalling that $\tau = \gamma/(\beta + \gamma)$, we can simply assume that $\beta$ is fixed. Therefore, for simplicity, we set the $\alpha$ and $\beta$ values to 2 and 1, respectively[4]. The change in punctuality requirements $\tau$ is due to the change in $\gamma$. Then, if the punctuality requirements $\tau$ are given, the excess travel time and travel time variability ratio can be easily computed according to Eqs. (11) and (24). Figure 8 shows the excess travel time and travel time variability ratio under different punctuality requirements $\tau$. With increasing $\tau$, the excess travel time increases but the travel time variability ratio decreases. This intuitively demonstrates the phenomenon of diminishing VOV given in Proposition 5 and Corollary 1. In other words, the different $\tau$ requirements of different travelers reflect the trade-off between excess travel time and the travel time variability ratio, i.e. the willingness of the traveler to accept an increase in travel time with certainty in exchange for a reduction in travel time risk. Specifically, as the traveler increases excess travel time (ETT) to mitigate for distribution tail and thereby increase the probability of arriving on-time, the value of further reducing travel time variability diminishes.

---
[4] $\alpha$ and $\beta$ are exactly the same as those in Small (1982), Fosgerau and Karlström (2010) and Taylor (2017).

Although $\tau$ is continuous in Figure 8, it would seem conceivable that in a realistic travel choice decision, the traveler might perceive the punctuality requirement in discrete increments, such as 5%. Based on Figure 8, we can obtain the excess travel time and travel time variability ratio for every 5% increase of $\tau$, as shown in Table 5. For example, a traveler with a current $\tau$ of 60% wishes to increase the punctuality requirement but is unwilling to accept that the value of the additional time he/she paid for hedging against travel time variability (i.e., the value of travel time variability, VOV) is diminished by more than 15%. In other words, the reduction of VOV should be less than 15%. Table 5 lists the proportion of change in the excess travel time relative to the initial excess travel time, i.e., 12.35, at $\tau = 0.60$ and the proportion of change in the travel time variability ratio relative to the initial travel time variability ratio, i.e., 0.6889, at $\tau = 0.60$ in parentheses. According to Table 5, if the traveler increases his/her $\tau$ to 0.70, his/her excess travel time would be 15.80, and thus the proportion of change in the excess travel time is $(15.80 − 12.35)/12.35 = 27.93\%$. As the travel time variability ratio for $\tau = 0.70$ is 0.65, the proportion of change in the travel time variability ratio is $(0.6457 − 0.6889)/0.6889 = −6.27\%$. In other words, the traveler faces a VOV reduction of 6.27% if he/she wishes to increase the $\tau$ from 0.6 to 0.7. If the traveler is unwilling to 'pay' for a VOV reduction greater than 15%, then a $\tau$ of 0.85 represents the best choice according to Table 5. Therefore, the travel time variability ratio can help travelers to achieve a better balance between trip reliability and trip cost.

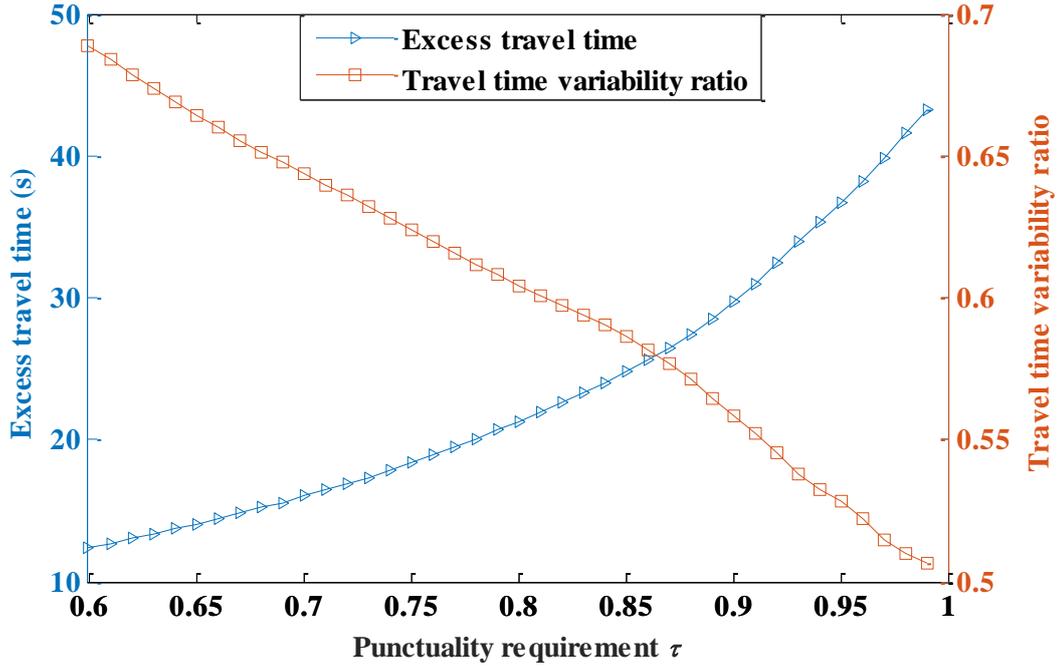

Figure 8. The excess travel time and the travel time variability ratio for different punctuality requirements

Table 5. Excess travel times and travel time variability ratios under different $\tau$.

| $T$ | Excess travel time | Travel time variability ratio |
|---|---|---|
| 0.60 | 12.35 | 0.6889 |
| 0.65 | 14.07 (↑13.92%) | 0.6642 (↓3.59%) |
| 0.70 | 15.80 (↑27.93%) | 0.6457 (↓6.27%) |
| 0.75 | 18.10 (↑46.55%) | 0.6256 (↓9.12%) |
| 0.80 | 20.99 (↑69.95%) | 0.6061 (↓12.02%) |
| **0.85** | 24.40 (↑97.56%) | 0.5884 (↓**14.59%**) |
| 0.90 | 29.14 (↑135.93%) | 0.5613 (↓18.52%) |

Note: Proportions of change relative to a punctuality requirement of 0.60 are indicated in parentheses. ↑ indicates an increase and ↓ indicates a decrease.

## 4. CONCLUSIONS

Within the framework of the standard Small scheduling model, the existing literature on valuing travel time variability has paid limited attention to the unexpected delay due to the unreliable aspect (i.e., distribution tail) of travel time. This is despite extensive evidence that unexpected delay can have much more serious consequences than expected or modest delay. The conceptual contribution of this paper was to propose the value of travel time *distribution tail* (VODT), capturing the unexpected delay inherent within the long fat tails that typically characterize travel time distributions. Having made this important definition, the paper reconciled the VODT with two existing concepts in the literature, namely the reliability premium (Batley, 2007) and mean-excess travel time (Chen and Zhou, 2010). Furthermore, the paper demonstrated that the aforementioned two concepts are in essence equivalent, with both concepts considering the increase in certain travel time that the traveler is willing to accept in exchange for reducing or indeed eliminating travel time variability.

The paper then introduced the more general concept of the value of travel time variability (VOV), as a weighted sum of the value of travel time *reliability* (VOR) and VODT and investigated the properties of excess travel time and the VOV, especially in relation to the scheduling parameters of the Small model. This analysis exposed the key property that, as the traveler increases excess travel time to mitigate for unreliability aspect of travel time and thereby increase the probability of arriving on-time, the marginal benefit of further reducing travel time variability diminishes. Based on the theoretically-proven diminishing marginal effect of the VOV under a validity condition, the paper further proposed the travel time variability ratio – an extension of the established reliability ratio (e.g., Jackson and Jucker, 1981) – as a means of helping travelers to understand the trade-off between punctuality requirements and the cost of travel time variability, and thereby determine appropriate punctuality requirements. We used empirical datasets to test the validity condition of the diminishing marginal effect of the VOV, and our results indicated that it is a relatively slack condition that can be easily satisfied. Besides, we presented a numerical example to illustrate how the travel time variability ratio can support a traveler's optimization of their punctuality requirements. Using numerical examples, we demonstrated that the distribution tail cost may account for more than 10% of the total cost of travel time variability. Therefore, if appraisals fail to consider the distribution tail of travel time and its impact on departure time and route choices, then this could seriously bias policy decisions.

In summary, the VODT can complement existing research on travel time variability by providing a more complete and definitive consideration of the cost of travel time variability. Specifically, the VOR quantifies the value of expected risk/duration of delay, while the VODT quantifies the value of unexpected risk/duration of delay. Therefore, the VODT can help travelers to better understand the value of mitigating for the more serious delay occurrences associated with the long fat tail of the travel time distribution, and help planners and policymakers to better understand the social costs and benefits of investments designed to reduce such delays. Besides, travelers may need to set their risk parameters in many emerging personalized mobility services such as routing navigation. The diminishing marginal effect of VOV suggests that it may be economically inefficient to blindly pursue a higher probability of not being late. That is to say, the proposed travel time variability ratio intuitively quantifies the implicit cost of the punctuality requirements that travelers impose for their trips, thereby supporting travelers in achieving a better balance between trip reliability and trip cost.

Based on our work, further research is warranted. The assumption that the standardized travel time is independent of the departure time could be relaxed to consider the time-varying travel time distribution. Furthermore, the step utility function used in this paper represents a special case of the utility function in the schedule delay model and implies a specific form of risk aversion, Eeckhoudt (2012) and other related references (Mas-Colell *et al*., 1995; Eeckhoudt and Schlesinger, 2006; Beaud *et al*., 2016; Eeckhoudt *et al*., 2022; Li *et al*., 2022) highlight that higher order derivatives may provide insight on the intensities of attitudes beyond risk aversion such as absolute prudence and absolute temperance. In a similar vein, we note Beaud (2016) works on the relationship between risk aversion and travel time and inferences regarding constant absolute risk aversion, increasing absolute risk aversion, and decreasing absolute risk aversion behaviors. Therefore, an investigation of the value of travel time distribution tail under other utility functions, such as the slope or nonlinear utility function (Vickrey, 1973; Tseng and Verhoef, 2008; Li *et al*., 2012), may be worthwhile. Besides, it would be valuable to extend the analysis of this paper from single trips to trip chains or path level (Jenelius *et al*., 2011; Jenelius, 2012, Jiang *et al*., 2022) and/or the network level (Uchida, 2014; Kato *et al*., 2021). Finally, given that information on travel time variability may influence travelers' choices of departure time (de Palma and Picard, 2006; de Palma *et al*., 2012; Lindsey *et al*, 2014; Engelson and Fosgerau, 2020, Jiang *et al*., 2020), it would be interesting to further explore the effect of the accuracy of the available information and the regimes of information release (de Palma *et al*., 2012; Han *et al*., 2021; Chen *et al*., 2023).

## ACKNOWLEDGMENT

This study was jointly sponsored by the National Natural Science Foundation of China (71971159 and 72021002), and the Fundamental Research Funds for the Central Universities (2022-5-YB-02). These supports are gratefully acknowledged.

# APPENDIX A. THE DEDUCTION OF EQUATION (2) FROM EQUATION (1)

As $T = A - D$, $PAT - A = (PAT - D) - (A - D)$ and $A - PAT = (A - D) - (PAT - D)$, by expanding terms of Eq. (1), we have

$$U(D, T) = -\left\{\alpha(A-D) + \beta\left((PAT-D)-(A-D)\right)^+ + \gamma\left((A-D)-(PAT-D)\right)^+\right\} \quad \text{(A-1)}$$

Rearranging Eq. (A-1):

$$\begin{aligned}-U(D, T) &= (\alpha - \beta)(A-D)\big|_{(PAT-A)^+} + (\alpha + \gamma)(A-D)\big|_{(A-PAT)^+} \\ &\quad + \beta(PAT - D)\big|_{(PAT-A)^+} - \gamma(PAT - D)\big|_{(A-PAT)^+}\end{aligned} \quad \text{(A-2)}$$

To simplify mathematical deductions, let $PAT = 0$. Since the travel time to arrive at exactly the PAT is $T = PAT - D = -D$, travel time must by definition be positive, this explains why $D < 0$ by the assumption $PAT = 0$. Then, Eq. (A-2) can be simplified as

$$\begin{aligned}&-U(D, T) \\ &= (\alpha-\beta)(A-D)\big|_{(PAT-A)^+} + (\alpha+\gamma)(A-D)\big|_{(A-PAT)^+} - \beta D\big|_{(PAT-A)^+} + \gamma D\big|_{(A-PAT)^+} \\ &= (\alpha-\beta)(A-D)\big|_{A<0} + (\alpha+\gamma)(A-D)\big|_{A\geq 0} - \beta D\big|_{A<0} + \gamma D\big|_{A\geq 0}\end{aligned} \quad \text{(A-3)}$$

Since $(\alpha - \beta)(A - D) = (\alpha - \beta)(A - D)\big|_{A<0} + (\alpha - \beta)(A - D)\big|_{A\geq 0}$, we have

$$\begin{aligned}-U(D, T) &= (\alpha-\beta)(A-D)\big|_{A<0} + (\alpha+\gamma)(A-D)\big|_{A\geq 0} - \beta D\big|_{A<0} + \gamma D\big|_{A\geq 0} \\ &= (\alpha-\beta)(A-D) + (\alpha+\gamma-(\alpha-\beta))(A-D)\big|_{A\geq 0} - \beta D + (\beta+\gamma)D\big|_{A\geq 0} \\ &= (\alpha-\beta)T + (\beta+\gamma)(A-D)\big|_{A\geq 0} - \beta D + (\beta+\gamma)D\big|_{A\geq 0} \\ &= (\alpha-\beta)T + (\beta+\gamma)(A-D+D)\big|_{A\geq 0} - \beta D \\ &= (\alpha-\beta)T + (\beta+\gamma)(T+D)\big|_{A\geq 0} - \beta D \quad (\text{Since } A = T + D)\end{aligned} \quad \text{(A-4)}$$

The final identity of Eq. (A-4) is essentially Eq. (2) in Section 2.1.

# APPENDIX B. PROOFS OF PROPOSITIONS AND LEMMAS

**Proof of Proposition 1**. The expected utility of a trip under uncertainty for a given departure time $D$ is $EU$, given by Eq. (4). According to Batley (2007), calculation of the reliability premium depends on identifying the certainty-equivalent trip that corresponds to the trip under uncertainty. A risk-averse traveler would choose to add the reliability premium to ensure that their certainty-equivalent arrival time is earlier than the PAT. Hence, for a given $D$, the trip utility of a certainty-equivalent arrival time before the PAT is

$$\begin{aligned}U(\mu + \pi) &= -\left\{(\alpha-\beta)(E(A)-(D-\pi)) - \beta(D-\pi)\right\} \\ &= -\left\{(\alpha-\beta)(\mu+\pi) - \beta D + \beta\pi\right\} \\ &= -\left\{(\alpha-\beta)\mu - \beta D + \alpha\pi\right\}\end{aligned} \quad \text{(B-1)}$$

where $E(A)$ is the certainty-equivalent arrival time. As $|EU| = |U(\mu + \pi)|$ when calculating the reliability premium $\pi$, $\pi$ can be obtained as

$$\pi = \frac{|EU| - (\alpha - \beta)\mu + \beta D}{\alpha} = \frac{\beta + \gamma}{\alpha}\left(\int_{\frac{-D-\mu}{\sigma}}^{\infty}(\mu + \sigma x + D) \cdot f(x)dx\right) \tag{B-2}$$

This completes the proof. □

**Proof of Proposition 3.** If the departure time $D$ in the reliability premium is given by Eq. (5), i.e., $D = -b(\tau) = -\mu - \sigma F^{-1}(\gamma/(\beta+\gamma))$, then the reliability premium with Eq. (B-2) will be:

$$\begin{aligned}
\pi &= \frac{\beta + \gamma}{\alpha}\left(\int_{\frac{-D-\mu}{\sigma}}^{\infty}(\mu + \sigma x + D) \cdot f(x)dx\right) \\
&= \sigma \cdot \frac{\beta + \gamma}{\alpha}\left(\int_{F^{-1}\left(\frac{\gamma}{\beta+\gamma}\right)}^{\infty}\left(x - F^{-1}\left(\frac{\gamma}{\beta+\gamma}\right)\right) \cdot f(x)dx\right) \\
&= \sigma \cdot \frac{\beta + \gamma}{\alpha}\int_{\frac{\gamma}{\beta+\gamma}}^{1}\left(F^{-1}(x) - F^{-1}\left(\frac{\gamma}{\beta+\gamma}\right)\right)dx = \frac{\beta + \gamma}{\alpha}S_u
\end{aligned} \tag{B-3}$$

As $\tau = \gamma/(\beta + \gamma)$, the expected excess delay is similarly restated as

$$\delta_{EED} = \sigma \cdot \frac{\beta + \gamma}{\beta}\int_{\frac{\gamma}{\beta+\gamma}}^{1}\left(F^{-1}(x) - F^{-1}\left(\frac{\gamma}{\beta+\gamma}\right)\right)dx = \frac{\beta + \gamma}{\beta}S_u \tag{B-4}$$

Therefore, $\delta_{EED} = (\alpha/\beta)\pi$. This completes the proof. □

**Proof of Proposition 4.** The first-order derivative of Eq. (4) is

$$\begin{aligned}
\frac{\partial EU(D,T)}{\partial D} &= -\left\{-\beta - (\beta + \gamma)(\mu - D - \mu + D)f\left(\frac{-D-\mu}{\sigma}\right)\frac{1}{\sigma} + (\beta + \gamma)\int_{\frac{-D-\mu}{\sigma}}^{\infty}f(x)dx\right\} \\
&= -\left\{-\beta + (\beta + \gamma)\int_{\frac{-D-\mu}{\sigma}}^{\infty}f(x)dx\right\}
\end{aligned} \tag{B-5}$$

Since $\partial EU(D,T)/\partial D = 0$ when $D = D^*$, according to Eq. (B-5), if $\partial EU(D,T)/\partial D > 0$, then $D < D^*$; if $\partial EU(D,T)/\partial D < 0$, then $D > D^*$. Consequently, we can infer that $EU(D, T)$ is strictly decreasing when $D > D^*$ and strictly increasing when $D < D^*$, with a maximum value at $D = D^*$. Note that the travel time budget associated with the punctuality requirement $\tau$ is exactly the optimal departure time, i.e., $D^* = D_{TTB}$. We can easily prove that the METT is greater than or equal to the TTB, as the EED is non-negative by definition:

$$\delta_{EED}(\tau) \geq 0, \forall \tau \geq \frac{\gamma}{\beta+\gamma} \Rightarrow \eta(\tau) = b(\tau) + \delta_{EED}(\tau) \geq b(\tau) \tag{B-6}$$

According to Eq. (B-6), $D_{METT} \leq D_{TTB}$, yielding $EU(D_{METT}, T) \leq EU(D_{TTB}, T)$ and thus $|EU(D_{METT}, T)| \geq |EU(D_{TTB}, T)|$. This completes the proof. $\square$

**Proof of Lemma 1.** The derivative of the ETT scaling factor $\zeta_{ETT}$ with respect to $\beta$ is

$$\begin{aligned}\frac{\partial \zeta_{ETT}}{\partial \beta} &= \frac{-\gamma}{\beta^2} \int_{\gamma/\beta+\gamma}^{1} F^{-1}(x)dx + \frac{\beta+\gamma}{\beta}\left(-F^{-1}\left(\gamma/\beta+\gamma\right)\right)\frac{-\gamma}{(\beta+\gamma)^2} \\ &= \frac{-\gamma}{\beta^2} \int_{\gamma/\beta+\gamma}^{1} F^{-1}(x)dx + \frac{\gamma}{\beta(\beta+\gamma)} F^{-1}\left(\gamma/\beta+\gamma\right) \\ &= -\frac{\gamma}{\beta^2}\left(\int_{\gamma/\beta+\gamma}^{1} F^{-1}(x) - F^{-1}\left(\gamma/\beta+\gamma\right)dx\right) \leq 0\end{aligned} \tag{B-7}$$

The derivative of the ETT scaling factor $\zeta_{ETT}$ with respect to $\gamma$ is

$$\begin{aligned}\frac{\partial \zeta_{ETT}}{\partial \gamma} &= \frac{1}{\beta}\int_{\gamma/\beta+\gamma}^{1} F^{-1}(x)dx + \frac{\beta+\gamma}{\beta}\cdot\left(-F^{-1}\left(\gamma/\beta+\gamma\right)\right)\frac{\beta}{(\beta+\gamma)^2} \\ &= \frac{1}{\beta}\int_{\gamma/\beta+\gamma}^{1} F^{-1}(x)dx - F^{-1}\left(\gamma/\beta+\gamma\right)\frac{1}{\beta+\gamma} \\ &= \frac{1}{\beta}\int_{\gamma/\beta+\gamma}^{1}\left(F^{-1}(x) - F^{-1}\left(\gamma/\beta+\gamma\right)\right)dx \geq 0\end{aligned} \tag{B-8}$$

Hence, $\zeta_{ETT}$ is monotonically non-increasing with respect to $\beta$ and non-decreasing with respect to $\gamma$. This completes the proof. $\square$

**Proof of Proposition 5.** (1) The derivative of the VOV with respect to $\beta$ is

$$\begin{aligned}\frac{\partial VOV}{\partial \beta} &= F(\zeta_{ETT}) + (\beta+\gamma)F(\zeta_{ETT})'\frac{\partial \zeta_{ETT}}{\partial \beta} + \frac{\int_{F(\zeta_{ETT})}^{1} F^{-1}(x)dx}{\zeta_{ETT}} \\ &\quad + \frac{\partial \zeta_{ETT}}{\partial \beta}(\beta+\gamma)\frac{-\zeta_{ETT}^2 F(\zeta_{ETT})' - \int_{F(\zeta_{ETT})}^{1} F^{-1}(x)dx}{\zeta_{ETT}^2} \\ &= F(\zeta_{ETT}) + \frac{\int_{F(\zeta_{ETT})}^{1} F^{-1}(x)dx}{\zeta_{ETT}} - \frac{\partial \zeta_{ETT}}{\partial \beta}(\beta+\gamma)\frac{\int_{F(\zeta_{ETT})}^{1} F^{-1}(x)dx}{\zeta_{ETT}^2}\end{aligned} \tag{B-9}$$

As $\frac{\partial \zeta_{ETT}}{\partial \beta} \leq 0$, $F(\zeta_{ETT}) \geq 0$, and $F^{-1}(\tau) \geq 0$ for risk-averse travelers, $\frac{\partial VOV}{\partial \beta} \geq 0$.

(2) Please refer to Appendix C for a detailed deduction of the complicated derivative of the VOV with respect to $\gamma$. The final result is

$$\frac{\partial VOV}{\partial \gamma} = -1 + F(\zeta_{ETT}) + \frac{F^{-1}\left(\gamma/\beta+\gamma\right)\int_{F(\zeta_{ETT})}^{1} F^{-1}(x)dx}{\zeta_{ETT}^2}$$

If $l \leq \kappa + 1$, then we obtain the following inequality:

$$\int_{F(\zeta_{ETT})}^{1} F^{-1}(x) dx \bigg/ \zeta_{ETT} \left(1 - F(\zeta_{ETT})\right) \leq \zeta_{EED} \bigg/ F^{-1}(\tau) + 1$$

$$\Rightarrow F^{-1}\left(\frac{\gamma}{\beta+\gamma}\right) \int_{F(\zeta_{ETT})}^{1} F^{-1}(x) dx \leq \zeta_{ETT}^{2} \left(1 - F(\zeta_{ETT})\right)$$

$$\Rightarrow -1 + F(\zeta_{ETT}) + \frac{F^{-1}\left(\frac{\gamma}{\beta+\gamma}\right) \int_{F(\zeta_{ETT})}^{1} F^{-1}(x) dx}{\zeta_{ETT}^{2}} \leq 0 \quad \text{(B-10)}$$

$$\Rightarrow \frac{\partial VOV}{\partial \gamma} \leq 0$$

Finally, the VOV is strictly non-increasing with respect to $\gamma$ for late arrival. This completes the proof. □

**Proof of Proposition 6.** (1) The derivatives of $\rho_{TTRR}$ and $\rho_{TTVR}$ with respect to $\alpha$ are, respectively:

$$\frac{\partial \rho_{TTRR}}{\partial \alpha} = -\frac{\beta+\gamma}{\alpha^{2}} \int_{\frac{\gamma}{\beta+\gamma}}^{1} F^{-1}(p) dp < 0$$

$$\frac{\partial \rho_{TTVR}}{\partial \alpha} = \frac{\gamma}{\alpha^{2}} - \frac{\beta+\gamma}{\alpha^{2}} F(\zeta_{ETT}) - \frac{\beta+\gamma}{\alpha^{2} \zeta_{ETT}} \int_{F(\zeta_{ETT})}^{1} F^{-1}(x) dx \quad \text{(B-11)}$$

$$\leq \frac{\gamma}{\alpha^{2}} - \frac{\beta+\gamma}{\alpha^{2}} F(\zeta_{ETT}) - \frac{\beta+\gamma}{\alpha^{2}} \left(1 - F(\zeta_{ETT})\right) = -\frac{\beta}{\alpha^{2}}$$

Obviously, $\frac{\partial \rho_{TTRR}}{\partial \alpha} < 0$ and $\frac{\partial \rho_{TTVR}}{\partial \alpha} < 0$.

(2) The derivatives of $\rho_{TTRR}$ and $\rho_{TTVR}$ with respect to $\beta$ are, respectively:

$$\frac{\partial \rho_{TTRR}}{\partial \beta} = \frac{1}{\alpha} \int_{\frac{\gamma}{\beta+\gamma}}^{1} \left(F^{-1}(p) - F^{-1}(\frac{\gamma}{\beta+\gamma})\right) dp \geq 0, \quad \frac{\partial \rho_{TTVR}}{\partial \beta} = \frac{1}{\alpha} \frac{\partial VOV}{\partial \beta} \geq 0 \quad \text{(B-12)}$$

(3) The derivatives of $\rho_{TTRR}$ and $\rho_{TTVR}$ with respect to $\gamma$ are, respectively:

$$\frac{\partial \rho_{TTRR}}{\partial \gamma} = \frac{1}{\alpha} \int_{\frac{\gamma}{\beta+\gamma}}^{1} F^{-1}(p) dp - \frac{\beta+\gamma}{\alpha} F^{-1}(\frac{\gamma}{\beta+\gamma}) \frac{\beta}{(\beta+\gamma)^{2}}$$

$$= \frac{1}{\alpha} \int_{\frac{\gamma}{\beta+\gamma}}^{1} \left(F^{-1}(p) - F^{-1}(\frac{\gamma}{\beta+\gamma})\right) dp \geq 0 \quad \text{(B-13)}$$

$$\frac{\partial \rho_{TTVR}}{\partial \gamma} = \frac{1}{\alpha} \frac{\partial VOV}{\partial \gamma}$$

Therefore, $\frac{\partial \rho_{TTVR}}{\partial \gamma} \leq 0$ if $l \leq \kappa + 1$. This completes the proof. □

## APPENDIX C. DERIVATIVE OF THE VALUE OF TRAVEL TIME VARIABILITY WITH RESPECT TO $\gamma$

$$\begin{aligned}\frac{\partial VOV}{\partial \gamma} &= -1 + F(\zeta_{ETT}) + f(\zeta_{ETT})\frac{\partial \zeta_{ETT}}{\partial \gamma}(\beta+\gamma) + \frac{\int_{F(\zeta_{ETT})}^{1} F^{-1}(x)dx}{\zeta_{ETT}} + (\beta+\gamma)\frac{-\zeta_{ETT}^{2} f(\zeta_{ETT}) - \int_{F(\zeta_{ETT})}^{1} F^{-1}(x)dx}{\zeta_{ETT}^{2}}\frac{\partial \zeta_{ETT}}{\partial \gamma}\\ &= -1 + F(\zeta_{ETT}) + \frac{\int_{F(\zeta_{ETT})}^{1} F^{-1}(x)dx}{\zeta_{ETT}} - (\beta+\gamma)\frac{\int_{F(\zeta_{ETT})}^{1} F^{-1}(x)dx}{\zeta_{ETT}^{2}}\frac{1}{\beta}\int_{\gamma/\beta+\gamma}^{1}\left(F^{-1}(x) - F^{-1}\left(\gamma/\beta+\gamma\right)\right)dx\\ &= -1 + F(\zeta_{ETT}) + \frac{\int_{F(\zeta_{ETT})}^{1} F^{-1}(x)dx}{\zeta_{ETT}}\left(1 - \frac{\int_{\gamma/\beta+\gamma}^{1} F^{-1}(x)dx - \frac{\beta}{\beta+\gamma}F^{-1}\left(\gamma/\beta+\gamma\right)}{\int_{\gamma/\beta+\gamma}^{1} F^{-1}(x)dx}\right)\\ &= -1 + F(\zeta_{ETT}) + \frac{\int_{F(t)}^{1} F^{-1}(x)dx}{\zeta_{ETT}}\frac{F^{-1}\left(\gamma/\beta+\gamma\right)}{\frac{\beta+\gamma}{\beta}\int_{\gamma/\beta+\gamma}^{1} F^{-1}(x)dx}\\ &= -1 + F(\zeta_{ETT}) + \frac{F^{-1}\left(\gamma/\beta+\gamma\right)\int_{F(\zeta_{ETT})}^{1} F^{-1}(x)dx}{\zeta_{ETT}^{2}}\end{aligned} \qquad (C\text{-}1)$$

## APPENDIX D. PROOF OF THE TRUTH OF $l \leq \kappa + 1$ IN THE REMARK 2

Note that $\zeta_{ETT} = F^{-1}(\tau) + \zeta_{EED}$, and $F(x)$ and $F^{-1}(x)$ are all non-decreasing functions. It is easy to have $F(\zeta_{ETT}) > \tau$ and thus $F^{-1}(F(\zeta_{ETT})) > F^{-1}(\tau)$. The following deduction shown in Eq. (C-1) indicates that $l \leq \kappa + 1$ is equivalent to a non-increasing $g(p)$ for $p \in [\tau, 1)$, where

$$g(p) = \frac{\int_{p}^{1} F^{-1}(x)dx}{1-p} - F^{-1}(p).$$

$$\frac{\int_\tau^1 F^{-1}(x)dx}{1-\tau} - F^{-1}(\tau) \geq \frac{\int_{F(\zeta_{ETT})}^1 F^{-1}(x)dx}{1-F(\zeta_{ETT})} - F^{-1}(F(\zeta_{ETT}))$$

$$\Rightarrow \frac{\int_\tau^1 F^{-1}(x)dx}{F^{-1}(\tau)(1-\tau)} - 1 \geq \frac{1}{F^{-1}(\tau)}\left(\frac{\int_{F(\zeta_{ETT})}^1 F^{-1}(x)dx}{1-F(\zeta_{ETT})} - F^{-1}(F(\zeta_{ETT}))\right) \geq \frac{1}{F^{-1}(F(\zeta_{ETT}))}\left(\frac{\int_{F(\zeta_{ETT})}^1 F^{-1}(x)dx}{1-F(\zeta_{ETT})} - F^{-1}(F(\zeta_{ETT}))\right)$$

$$\Rightarrow \frac{\int_\tau^1 F^{-1}(x)dx}{F^{-1}(\tau)(1-\tau)} - 1 \geq \frac{1}{F^{-1}(F(\zeta_{ETT}))}\left(\frac{\int_{F(\zeta_{ETT})}^1 F^{-1}(x)dx}{1-F(\zeta_{ETT})} - F^{-1}(F(\zeta_{ETT}))\right) \quad \text{(D-1)}$$

$$\Rightarrow \frac{\int_\tau^1 F^{-1}(x)dx}{F^{-1}(\tau)(1-\tau)} - 1 \geq \frac{\int_{F(\zeta_{ETT})}^1 F^{-1}(x)dx}{F^{-1}(F(\zeta_{ETT}))(1-F(\zeta_{ETT}))} - 1$$

$$\Rightarrow \frac{F^{-1}(F(\zeta_{ETT}))}{F^{-1}(\tau)} \geq \frac{\int_{F(\zeta_{ETT})}^1 F^{-1}(x)dx}{1-F(\zeta_{ETT})} \cdot \frac{1}{\frac{\int_\tau^1 F^{-1}(x)dx}{(1-\tau)}} \Leftrightarrow \kappa+1 \geq l$$

The derivative of $g(p)$ with respect to $p$ is

$$g(p)' = \left(\frac{\int_p^1 F^{-1}(x)dx}{1-p} - F^{-1}(p)\right)' = \frac{-F^{-1}(p)(1-p) + \int_p^1 F^{-1}(x)dx}{(1-p)^2} - F^{-1}(p)' = \frac{-F^{-1}(p)(1-p) - F^{-1}(p)'(1-p)^2 + \int_p^1 F^{-1}(x)dx}{(1-p)^2} \quad \text{(D-2)}$$

As $(1-p)^2 > 0$, Eq. (D-2) indicates that the sign of $g(p)'$ is the same as the sign of $u(p)$, where $u(p)$ is the numerator of the last fraction in Eq.

(D-2), i.e., $u(p) = -F^{-1}(p)(1-p) - F^{-1}(p)'(1-p)^2 + \int_p^1 F^{-1}(x)dx$. The derivative of $u(p)$ with respect to $p$ is

$$\begin{aligned} u(p)' &= \left(-F^{-1}(p)(1-p) - F^{-1}(p)'(1-p)^2 + \int_p^1 F^{-1}(x)dx\right)' \\ &= -F^{-1}(p)'(1-p) + F^{-1}(p) - F^{-1}(p)''(1-p)^2 + 2F^{-1}(p)'(1-p) - F^{-1}(p) \\ &= F^{-1}(p)'(1-p) - F^{-1}(p)''(1-p)^2 \end{aligned}$$ (D-3)

Note that $g(p)' \leq 0$ requires that $u(p) \leq 0$. Considering $u(1) = 0$, if we have $u(p)' \geq 0$ for $p \in [\tau, 1)$, then we have $u(p) \leq 0$. According to Eq. (D-2), $u(p)' \geq 0$ when $\dfrac{F^{-1}(p)''}{F^{-1}(p)''} \leq \dfrac{1}{1-p}$. Consequently, $l \leq \kappa + 1$ is true as long as $\dfrac{F^{-1}(p)''}{F^{-1}(p)''} \leq \dfrac{1}{1-p}$. Specifically, if $F^{-1}(p)$ is a concave function, then $F^{-1}(p)'' \geq 0$ and thus $l \leq \kappa + 1$ must be true.

## NOTATION

| Notation | Definition |
|---|---|
| $T, X$ | Travel time and standardized travel time |
| $\varphi, \Phi$ | PDF and CDF of $T$ |
| $\mu, \sigma$ | Mean and standard deviation of $T$ |
| $f, F$ | PDF and CDF of $X$ |
| $\tau$ | Punctuality requirement (i.e. probability or confidence level of arriving at or before the Preferred Arrival Time (PAT)) |
| $\bar{T}, \bar{T}_{\min}$ | Travel time and minimal travel time required to ensure a desired $\tau$ |
| $b(\tau), \eta(\tau)$ | Travel time budget and mean-excess travel time for a desired $\tau$ |
| $\delta, \delta_{TTM}, \delta_{EED}$ | Safety margin, travel time margin, and expected excess delay |
| $\zeta_{TTM}, \zeta_{EED}, \zeta_{ETT}$ | Scaling factors of travel time margin, expected excess delay, and excess travel time |
| $\kappa, l$ | Quotient of $\zeta_{EED}$ divided by $F^{-1}(\tau)$ and quotient of $\int_{F(\zeta_{ETT})}^{1} F^{-1}(x)dx$ divided by $\zeta_{ETT}(1-F(\zeta_{ETT}))$ |
| $\alpha, \beta, \gamma$ | Scheduling preference parameters |
| $U, EU$ | Trip cost and expected trip cost in utility units |
| $A, D, D^*$ | Arrival time, departure time, and optimal departure time |